\documentclass[aps,prl,reprint,superscriptaddress,letterpaper,twocolumn,floatfix]{revtex4-1}

\usepackage{verbatim}
\usepackage{lipsum}
\usepackage{color}
\usepackage{amssymb}
\usepackage{amsmath,mathrsfs}
\usepackage{braket}
\usepackage{bm} 

\usepackage{setspace,soul}
\usepackage{graphicx}
\usepackage[dvipsnames]{xcolor}
\usepackage{grffile}
\usepackage{subfigure}

\usepackage[utf8]{inputenc}
\usepackage{geometry}
\usepackage{float}

\newcommand{\pot}{\Phi}
\newcommand{\pathV}{{\mathcal{V}}}

\newcommand{\rD}{\textcolor{red}{D}}
\newcommand{\rE}{\textcolor{red}{E}}

\newcommand{\bK}{\textcolor{blue}{K}}
\newcommand{\bR}{\textcolor{blue}{R}}
\newcommand{\bH}{\textcolor{blue}{H}}

\newcommand{\gC}{\textcolor{OliveGreen}{C}}

\usepackage{xparse}
\ExplSyntaxOn
\NewDocumentCommand{\mref}{m}{\quinn_mref:n {#1}}
\seq_new:N \l_quinn_mref_seq
\cs_new:Npn \quinn_mref:n #1
 {
  \seq_set_split:Nnn \l_quinn_mref_seq { , } { #1 }
  \seq_pop_right:NN \l_quinn_mref_seq \l_tmpa_tl
  ( 
  \seq_map_inline:Nn \l_quinn_mref_seq
    { \ref{##1},\nobreakspace } 
  \exp_args:NV \ref \l_tmpa_tl 
  ) 
 }
\ExplSyntaxOff

\usepackage{xr-hyper}
\usepackage[hyperindex,breaklinks]{hyperref}
\usepackage{filecontents}
\makeatletter\@input{xx.tex}\makeatother
\usepackage{pifont}

\begin{document}

\title{Mutational paths with sequence-based models of proteins:\\ from sampling to mean-field characterization}
\author{Eugenio Mauri, Simona Cocco, R\'emi Monasson}
\affiliation{Laboratory of Physics of the Ecole Normale Sup\'{e}rieure, CNRS UMR 8023 and PSL Research, Sorbonne Universit\'e, 24 rue Lhomond, 75231 Paris cedex 05, France}

\begin{abstract}
Identifying and characterizing mutational paths is an important issue in evolutionary biology, with potential applications to bioengineering. 
We here propose an algorithm to sample mutational paths, which we  benchmark on exactly solvable models of proteins in silico, and apply to data-driven models of natural proteins learned from  sequence data with Restricted Boltzmann Machines. We then use mean-field theory to characterize paths  for different mutational dynamics of interest, and to extend Kimura's estimate of evolutionary distances to sequence-based epistatic models of selection. 
\end{abstract}

\date{\today}

\maketitle

\paragraph{Introduction.}

Obtaining proteins with controlled properties, such as stability, binding affinity and specificity is a central goal in bioengineering \cite{DirectEvoArnold1997}. 
Over the past years, much progress on design was made using data-driven models, intended to capture the relation between protein sequences and  functionalities. In particular, unsupervised machine-learning approaches such as Boltzmann Machines (BM) or Variational Auto-Encoders trained on homologous sequence data (defining a protein family) were  able to design new proteins with functionalities comparable to natural proteins \cite{Russ2020Jul,Hawkins2021}. By comparison, the (even) harder problem of designing paths of sequences, interpolating between two homologous proteins has received little attention, see however~\cite{plosCBwest}. Yet solving this problem would shed light on the navigability of the sequence landscape \cite{navigability2021} and on how functional specificity, such as binding to distinct substrates could have emerged from ancestral, promiscuous proteins in the course of evolution~\cite{promiscuity}. In turn, it could help design new proteins interpolating between functional classes.


While various methods exist for building transition paths between the minima of a multi-dimensional continuous landscape \cite{vanden2010transition, bolhuis2002transition} dealing with discrete configurations requires the development of specific procedures \cite{mora2012transition}. We hereafter propose a Monte Carlo algorithm to sample mutational paths in protein landscapes, {\em e.g.} obtained by Restricted Boltzmann Machines trained on sequence data. We first benchmark our sampling procedure on an exactly solvable model of lattice proteins \cite{Lau1989Oct}, and demonstrate its capability to find high-quality paths between two proteins belonging to different subfamilies. We then apply our algorithm to the  WW domain, a binding module involved in the regulation of protein complexes \cite{WWstructure,Socolich2005}. The functionality of the sequences along the paths is validated with structure (ligand+protein)-informed software \cite{MPNN2022}. Last of all we derive a mean-field characterization of paths, tailored to the mutational dynamics of interest. This mean-field theory allows us to efficiently estimate evolutionary distances in the presence of strong epistatis in the selection process, which is not possible with profile models at the basis of most phylogenetic studies \cite{felsenstein2004inferring}. 

\paragraph{Definition and sampling of mutational paths.} We assume the sequence landscape  is modeled through a probability distribution $P_{model}({\bf v})$ over amino-acid sequences $\bf v$ of length $N$.  Informally speaking, $P_{model}$ quantifies the probability that $\bf v$ is a member of the protein family of interest, {\em i.e.} share its common structural and functional properties, and can be learned from homologous sequence data \cite{morcos2011direct,coccorev}. For natural protein families, exact expressions for $P_{model}$ are not available, but approximate distributions can be inferred from multi-sequence alignments (MSA) using unsupervised learning techniques. Previous works have shown that the inferred $P_{model}$ can serve as a proxy for the protein fitness \cite{shekhar2013spin,Jacquin_LP_2016, digiacchino, bisardi2022modeling}.

Hereafter, we use Restricted Boltzmann Machines (RBM)~\cite{fischer2012introduction}, a class of generative models based on two-layer graphs~\cite{tubiana_learning_2019}. RBM define a joint probability distribution of the protein sequence $\mathbf{v}$ (carried by the visible layer) and of its $M$-dimensional latent representation $\mathbf{h}$ (present on the hidden layer) as
\begin{equation}
P_{RBM} \propto \exp \left( \sum_{i=1}^N g_i(v_i) + \sum_{\mu=1}^M h_\mu I_\mu({\bf v}) - \sum_{\mu=1}^M \mathcal{U}_\mu(h_\mu) \right)\ ,
\label{eq:rbm}
\end{equation}
where  $I_\mu({\bf v}) = \sum_i w_{i,\mu}(v_i)$ is the input to hidden unit $\mu$. The $g_i$'s and $\mathcal{U}_\mu$'s are local potentials acting on, respectively, visible and hidden units, and the $w_{i\mu}$'s are the interactions between the two layers. They are learned by maximizing the marginal probabilities $P_{model}({\bf v}) = \int d{\bf h}\, P_{RBM}({\bf v},{\bf h})$ over the sequences $\bf v$ in a multi-sequence alignment of the family. While  other unsupervised procedures providing approximate $P_{model}$ can be used, such as Direct Coupling Analysis \cite{morcos2011direct,coccorev}, RBM offer a convenient way to monitor the changes in sequences along mutational paths, as we will see below.  

We consider mutational paths of $T$ steps, $\pathV=\{{\bf v}_1,{\bf v}_2,..., {\bf v}_{T-1}\}$, anchored at their extremities defined by the sequences $\mathbf{v}_{start}$ and $\mathbf{v}_{end}$.
The probability of a  path reads
\begin{eqnarray}
    & &\mathcal{P}[\pathV | \mathbf{v}_{start},\mathbf{v}_{end}] \propto   \prod_{t=1}^{T-1} P_{model}(\mathbf{v}_t) \times \nonumber \\ 
   & &  \pi(\mathbf{v}_{start},\mathbf{v}_1)   \times    \prod_{t=1}^{T-2}\pi(\mathbf{v}_t,\mathbf{v}_{t+1}) \times \pi(\mathbf{v}_{T-1},\mathbf{v}_{end})
    \label{eq:path_weight}
\end{eqnarray}
where the 'transition' factor $\pi(\mathbf{v},\mathbf{v}')$ increases with the similarity between the sequences $\mathbf{v},\mathbf{v}'$. In practice we choose $\pi=1$ if the two sequences are identical, $e^{-\Lambda}$ if they differ by one mutation (with $\Lambda>0 $), and $0$ if they are two or more mutations apart. This choice generates `continuous' paths, along which successive sequences differ by one mutation at most. Other choices for $\pi$, more plausible from an evolutionary point of view will be introduced below.

The probability ${\cal P}(\pathV)$ can be sampled as follows. Starting from a path $\pathV^{0}$, we randomly pick up an intermediate sequence ${\bf v}_t$ and attempt at mutating one amino acid, under the constraint that the Hamming distance of the trial sequence $\mathbf{v}'$ with $\mathbf{v}_{t-1}$ and $\mathbf{v}_{t+1}$ be at most 1. The mutation is then rejected or accepted, {\em i.e.} $\mathbf{v}_t \leftarrow \mathbf{v}'$ according to detailed balance. 
To improve the quality of the sampled mutational paths we introduce a fictitious inverse temperature $\beta$ and resort to simulated annealing. We then sample paths from ${\cal P}[\pathV]^\beta$, where $\beta$ is initially very small and is progressively ramped up to some target value. The complete procedure and the proof of detailed balance are given in Supplemental Material, Sec.~\ref{SI:path_alg}.

\begin{figure}
    \centering
    \includegraphics[width=\linewidth]{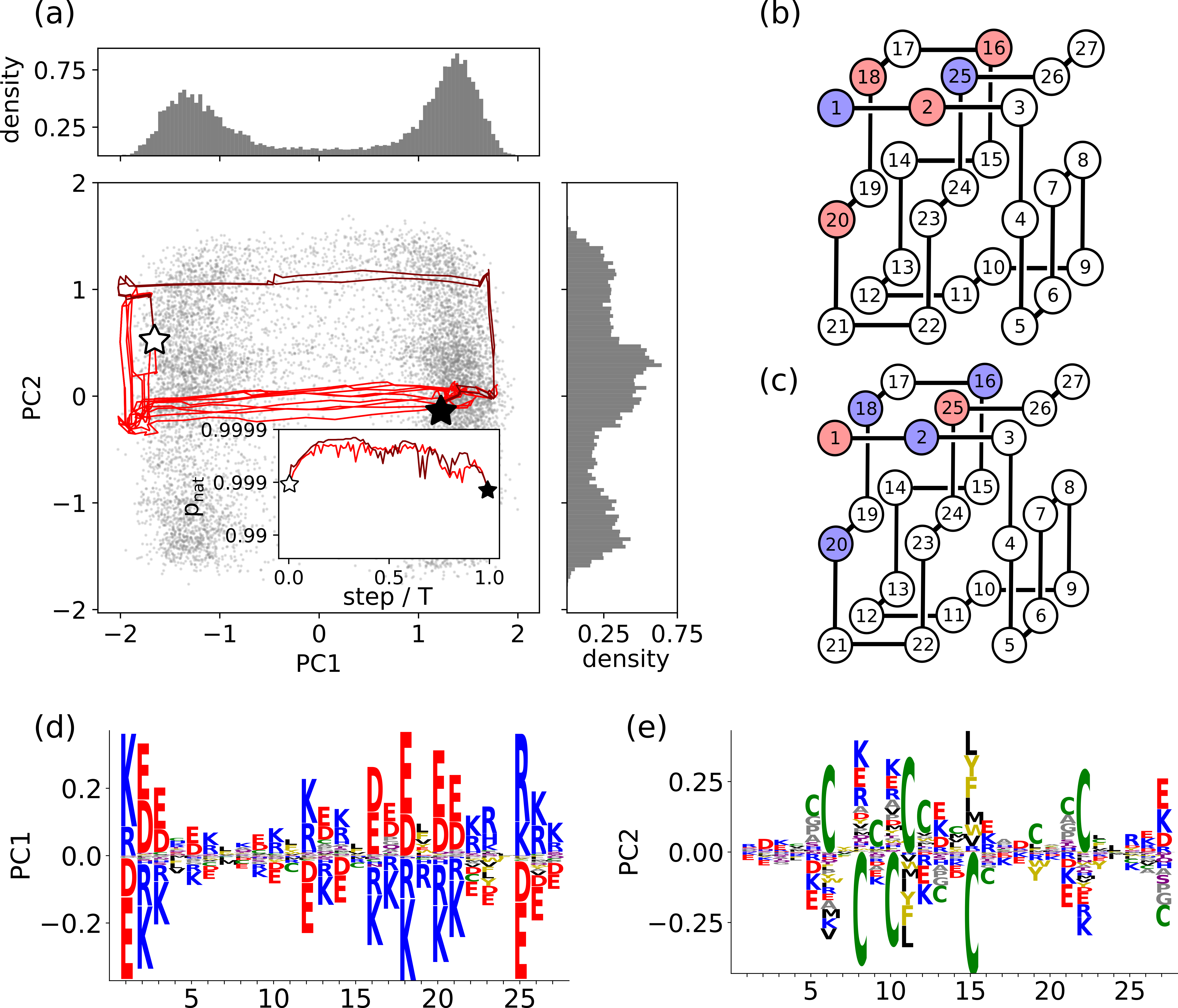}
    \caption{{\bf Mutational paths for lattice proteins}, joining sequences  \scalebox{1.2}{\ding{73}}=\texttt{{\rD\bR}GIQ{\gC}LAQMF{\rE\bK\rE}M{\bR\bK\bK\bR\bR\bK\gC}YL{\rE\gC\rD}} and \scalebox{1.2}{\ding{72}}=\texttt{{\bR\rE\gC\gC}AV{\gC\bH}Q{\bR}F{\bK\rD\bK}I{\rD\rE\rD}Y{\rE\rD}AWL{\bK\gC}N} belonging  to the family with structure shown in b) and c). Red and blue colors respectively correspond to negatively and positively charged amino acids. Cysteine is denoted by a green C. {(a)} Projections of $10^4$ LP sequences in the family (grey dots) along the top two PC of their correlation matrix. 
    Red and maroon lines show some representative paths sampled from Eq.~\eqref{eq:path_weight}. The relative numbers of maroon (2) and red (10) paths respect the statistics over all sampled paths. Parameter values: $\beta = 3$, $\Lambda=2$,  $T=82$.  Sides: histograms of projections along PC1 (top) and PC2 (right). Inset:  folding probabilities $p_{nat}$ along each path.(b,c) The fold of the LP family is stabilized by  alternating configurations of charges.  (d,e) Sequence logos of PC1 and PC2. }
    \label{fig:LP}
\end{figure}

\paragraph{Benchmarking mutational path sampling on in silico proteins.}
We benchmark the performances of our MC procedure on a model of Lattice Proteins (LP)~\cite{Lau1989Oct,shakh}. In LP, sequences of 27 amino acids may fold into $\simeq 10^5$ different self-avoiding conformations going through the nodes of a $3\times 3\times 3$ cubic lattice. The sequence landscape associated to a structure $\mathbf{S}$ (Fig.~\ref{fig:LP}(a)) is defined by the probability $p_{nat}(\mathbf{v}|\mathbf{S})$ that a sequence ${\bf v}$ has  $\mathbf{S}$ as its native fold; $p_{nat}$ can be exactly computed from the energies of interactions between adjacent amino acids, see Supplemental Material, Sec.~\ref{SI:LP} for details.

We first generate many sequences $\bf v$ with high $p_{nat}$ values for the fold $\bf S$ of Figs.~\ref{fig:LP}(b,c) following the procedure of \cite{Jacquin_LP_2016}. We next compute the top two Principal Components (PC) of these sequence data using one-hot encoding: PC1 corresponds to an extended electrostatic mode, and PC2 identifies possible Cys-Cys bridges (Figs.~\ref{fig:LP}(d,e)). Projecting the sequences onto these two PCs reveals two sub-families separated along PC1 (Fig.~\ref{fig:LP}(a)), associated to opposite chains of alternating charges along the electrostatic mode (Figs.~\ref{fig:LP}(b,c)). We will use our path sampling procedure to interpolate between the two sub-families, see start (white star) and end (black star) sequences in Fig.~\ref{fig:LP}(a).

To mimick the approach followed for natural proteins we train a RBM on the LP sequence data generated above, to infer an approximate expression for $p_{nat}$ from the data; see Supplemental Material, Sec.~\ref{SI:RBMtrain} for the inference of the RBM model. We then use our sampling algorithm to produce  mutational paths, see Fig.~\ref{fig:LP}(a). The algorithm is able to find excellent mutational paths in terms of the ground--truth  folding probabilities  $p_{nat}$ of intermediate sequences, even  higher than the ones of $\mathbf{v}_{start},\mathbf{v}_{end}$ when imposing high $\beta$  (insert of Fig.~\ref{fig:LP}(a)). Repeated runs of the sampling procedure give different paths that cluster into two classes, shown in red and maroon in Fig.~\ref{fig:LP}(a). While few  paths exploit a transient introduction of Cys-Cys interaction (on sites 6, 11 and 22) to stabilize the structure while flipping the electrostatic residues (maroon cluster); most introduce additional stabilizing electrostatic contacts along the path (red cluster). See Supplemental Material, Sec~\ref{SI:LP_RBM_stats} for details.

\paragraph{Mutational path sampling from data-driven models of natural proteins.}

\begin{figure}
    \centering
    \includegraphics[width=\linewidth]{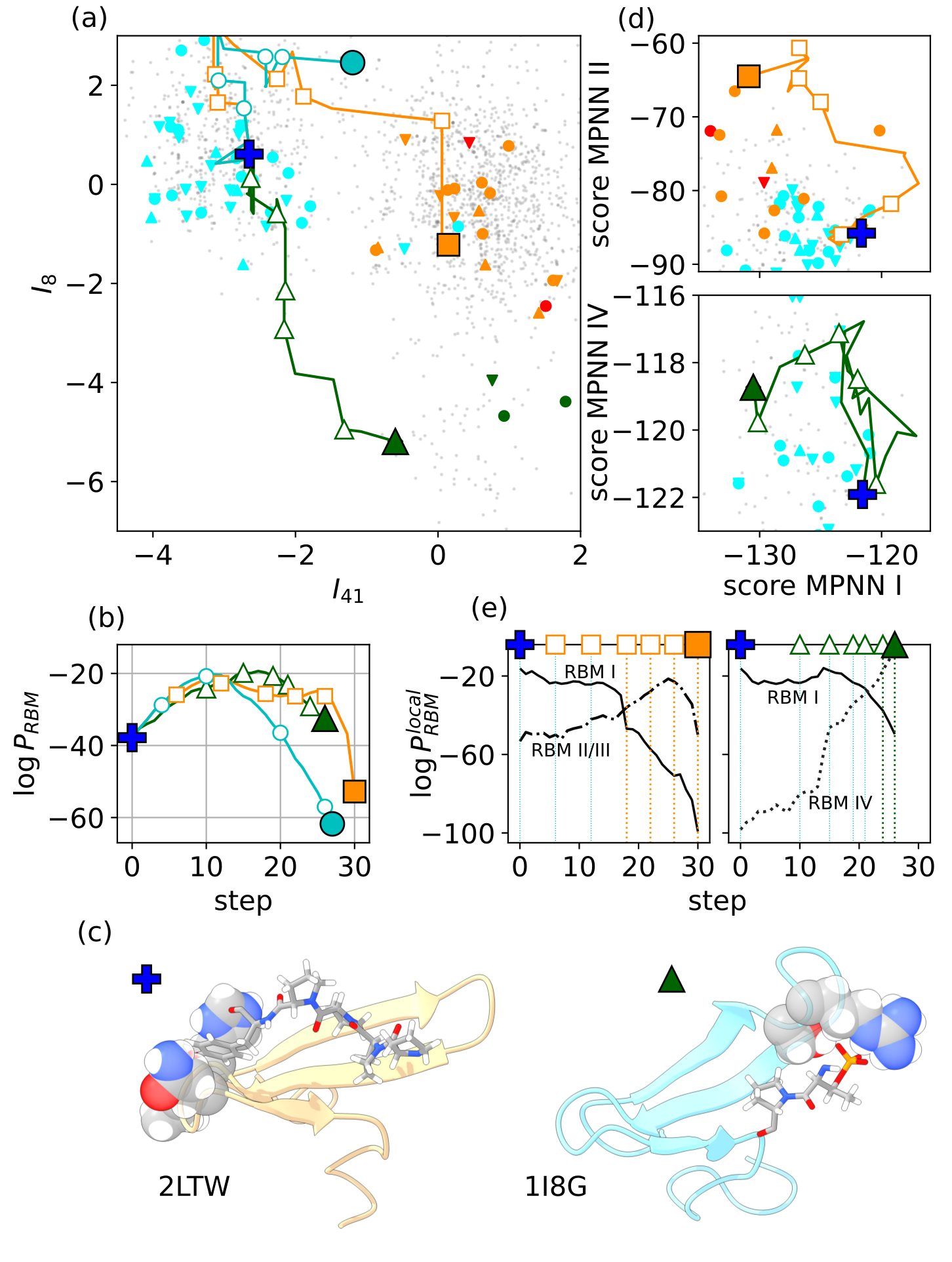}
    \caption{{\bf Mutational paths of the WW domain} using RBM  trained on the PFAM PF00397 family, see Supplemental Material, Sec.~\ref{SI:RBMtrain} for details about implementation.
    (a) Natural sequences $\bf v$ (grey dots) projected onto the plane of inputs $I_\mu$ (here $M=50$) of two hidden units clustering sequences according to the types of ligands they bind \cite{WWprolinerich}: I (cyan), II (red), III (orange), IV (green). Marked sequences: Upper (Lower) triangles: natural (artificial), from~\cite{russ2005natural}. Circles: natural, from~\cite{otte2003ww}. Blue cross represents the YAP1 domain.
    Lines shows the projection of three representative paths connecting YAP1 to sequences in classes I (circle),  II (square) and IV (triangle). Intermediate sequences (empty symbols) are listed in Supplemental Material, Sec.~\ref{SI:WW_list}. Parameters:  $\beta=3$, $\Lambda=0.1$.{(b)} Log $P_{RBM}$ for sequences along the paths. 
    (c) Complexes (WW domain and cognate peptides) for classes I (blue cross) and IV (green triangle)~\cite{chimerax}. Atoms corresponding to the two binding pockets are highlighted. 
    (d) ProteinMPNN scores for binding affinity, see Supplemental Material, Sec.~\ref{SI:MPNN}. $x$-axis measures the affinity to class I reference structure while $y$-axes show affinity to classes II/III (Top) and  IV (Bottom) reference structure respectively. 
    {(e)} Log-likelihood along the  paths from I to II (Left) and from I to IV (Right) according to class-specific RBM trained on  sequences in the three quadrants (Solid: I, Dot-dashed: II/III, Dotted: IV).
    }
    \label{fig:WW}
\end{figure}

We next show that our path sampling procedure can be applied to natural proteins. We train a RBM from MSA data of the WW family, a protein domain binding specifically proline-rich peptides \cite{WWprolinerich, WWstructure} and sample mutational paths between the Human YAP1 domain and three natural sequences known to have different binding specificities \cite{WWspecificities}. 
Figure~\ref{fig:WW}(a) shows some sampled paths in the plane spanned by the inputs $I({\bf v})$ (Eq. \ref{eq:rbm}) to two RBM hidden units chosen to cluster natural WW sequences depending on their binding specificities~\cite{tubiana_learning_2019}. Intermediate sequences have high probabilities according to the RBM model, see Fig.~\ref{fig:WW}(b). We then use AlphaFold~\cite{jumper2021highly} to assess the quality of the intermediates sequences; AlphaFold is able to predict the phenotypic effects of few mutations \cite{AF2_Tvsi}, and to compare the resulting structures to natural folds through Template Modelling scores ($\text{TM-score}$)~\cite{TMscore}, ranging from from 0 -unrelated proteins- up to 1 -perfect match. We obtain  $\text{TM-score}> 0.5$, indicating a high similarity between the folds of sequences sampled along the path and of natural WW, see Supplemental Material, Sec.~\ref{SI:WW_TM_score} for details.

We next estimate binding affinities for each class using ProteinMPNN \cite{MPNN2022}, an autoregressive structural-based probabilistic model that takes as input a backbone structure of a protein-ligand complex  and predicts the affinity score of a putative protein sequence. Here, we use available complexes of natural WW domains of binding classes I, II/III, IV with their cognate peptides, see Fig.~\ref{fig:WW}(c) and Supplemental Material, Sec.~\ref{SI:MPNN}. As expected, along the I $\to$ II/III path the affinities to class I (respectively, II/III)--cognate peptides decrease (increase), see Fig.~\ref{fig:WW}(d).
Interestingly, Fig.~\ref{fig:WW}(e) shows the existence of  a region on the I $\to$ IV path in which the predicted affinities with respect to both complexes  are high. It has been experimentally shown that some natural WW domains belonging to class I have also class IV activity~\cite{russ2005natural}. This promiscuity may be favored by the fact that class I and IV cognate peptides bind two distinct loops of the WW domain (Fig.~\ref{fig:WW}(c)). In Supplemental Material, Sec.~\ref{SI:ItoIVadditionalstatistics} we corroborate these results by sampling more paths between class I and IV. To further assess the specificity of sequences on the sampled path, we train approximate class-specific RBM models from sequences in the quadrants of Fig.~\ref{fig:WW}(a), see Supplemental Material, Sec.~\ref{SI:RBMtrain}. The cross-overs between the log-likelihoods of the class-specific RBMs in Fig.~\ref{fig:WW}(f,g) suggest the presence of specificity switches along the I $\to$ II/III and I $\to$ IV paths. The scores provided by class-specific RBMs and ProteinMPNN are correlated along the paths, see Supplemental Material Fig.~\ref{fig:SI_MPNN_RBM_compared}.

\paragraph{Mean-field theory of mutational paths.} To better characterize the typical properties of mutational paths  we resort to mean-field theory, by formally sending $N\to\infty$, while keeping the number $T$ of steps finite. To allow for  $\mathcal{O}(N)$ mutations between contiguous sequences we write the transition factor in Eq.~\ref{eq:path_weight} as
$\pi(\mathbf{v},\mathbf{v}') = e^{-N \Phi(q)}$, where the potential $\Phi$ is a decreasing function of the overlap $q=\frac 1N \sum_i \delta_{v_i,v'_i}$. $\Phi$ controls the elastic properties of the path, and will be made precise below.

Mean-field theory exploits the bipartite nature of the RBM architecture and allows us to monitor two sets of order parameters characterizing the paths $\pathV$: the mean values of the hidden-unit inputs, $m^\mu _t=\frac 1N \langle I_{\mu}(\mathbf{v}_t) \rangle$, and of the overlaps (fraction of conserved amino acids between successive sequences), $q_t= \frac 1N \sum_i \langle \delta_{v_{i,t},v_{i,t+1}}\rangle$; here, $\langle \cdot\rangle$ denotes the average over ${\cal P}(\pathV)^\beta$.

The $T\times (M+1)$ order parameters $m^\mu _t$ and $q_t$  are determined through minimization of the path free--energy density $f_{path}$, see Supplemental Material, Sec.~\ref{SI:RBM_MF_eqs}, with
\begin{align}
    &f_{\text{path}}(\{m^\mu_t\},\{q_t\}) = -\sum_{t,\mu}\left({\Gamma}_\mu(m_t^\mu)-m_t^\mu\,{\Gamma}_\mu'(m_t^\mu)\right)\\& +  \sum_t\left(\pot (q_t)-q_t\,\pot'(q_t)\right) - \frac{1}{\beta N}\sum_i\ln Z_i\big(\{m^\mu_t\},\{q_t\}\big)\ .\nonumber
    \label{eq:free_energy}
\end{align}
Here, $\Gamma_\mu(m)=\frac 1N \ln \int \mathrm{d} h \, e^{N\, m\, h-\mathcal{U}_\mu(h)}$ 
and $Z_i$ is the following site-dependent partition function,
\begin{multline}
    Z_i(\{m^\mu_t\},\{q_t\}) =  \sum_{\{v_t\}}\exp \left(\beta\sum_t g_i(v_{t}) \, +\right. \\ \left.+ \beta\sum_{t,\mu} {\Gamma}_\mu'(m_t^\mu)\, {w_{i\mu}(v_t)} -\beta\sum_t\pot'(q_t) \,\delta_{v_t,v_{t+1}} \right)\ . 
\end{multline}
$Z_i$ can be efficiently estimated through products of transfer matrices, of sizes $21\times 21$. While the expression of $f_{path}$ is exact for sequence length $N\to \infty$, we show below it is accurate 
even in the cases of LP ($N=27$) and WW ($N=31$).

\paragraph{Choice of the elastic potential.} The potential $\Phi$ can enforce   continuity (Cont) requirements, e.g. successive sequences along the path  differ by, say, $K$ mutations at most, or mimic the evolutionary (Evo) dynamics of natural sequences through stochastic mutations.

\begin{figure}
    \centering
    \includegraphics[width=\linewidth]{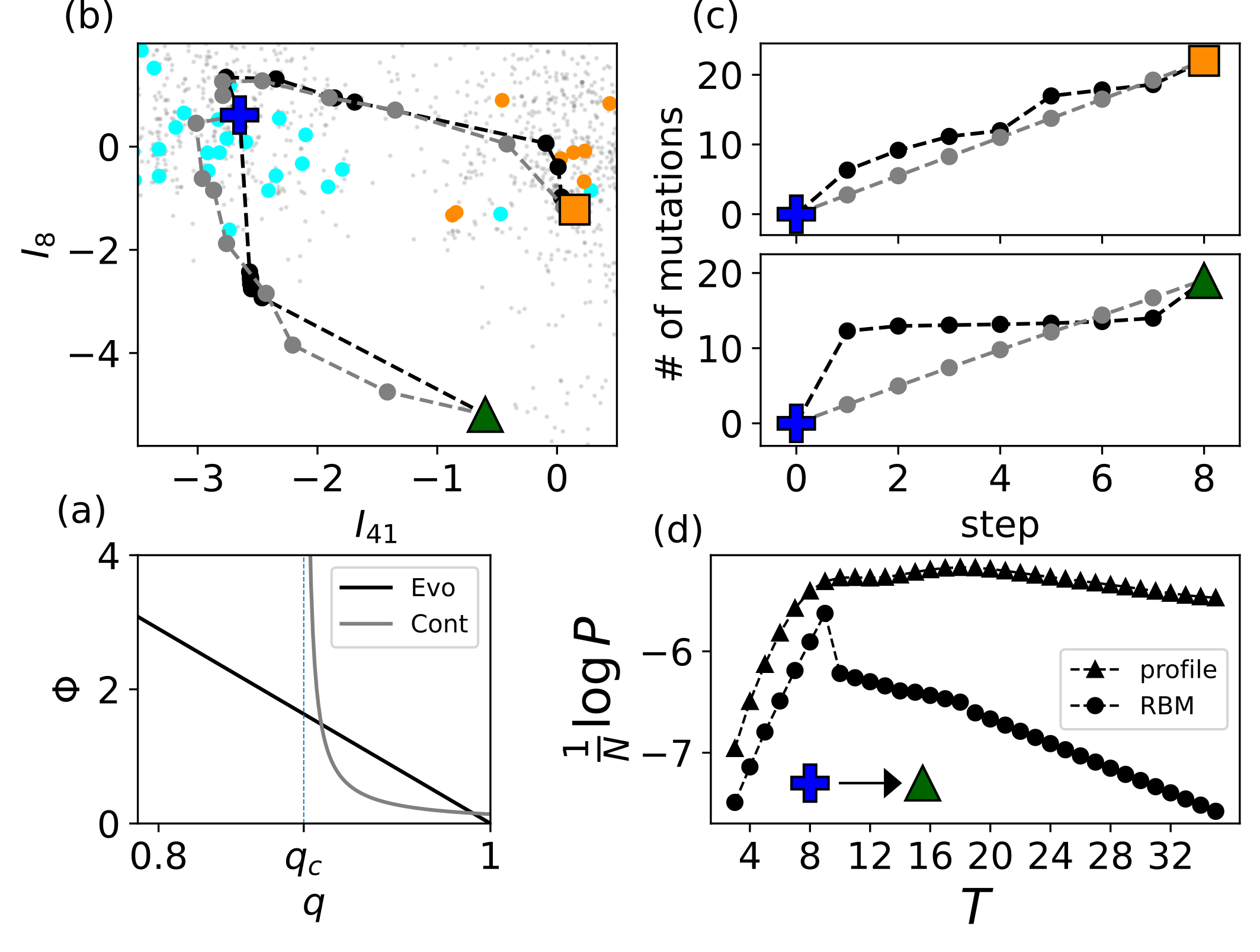}
    \caption{{\bf Mean-field theory of mutational paths} for the RBM model trained on WW domain. (a) Sketches of the potentials $\Phi_\text{Evo}$ (black) and  $\Phi_\text{Cont}$ (gray) vs. $q$.  (b) Same two-dimensional representation as in Fig.~\ref{fig:WW}(a) for the mean-field paths with Evo (black lines) and Cont (grey lines) potentials. (c) Cumulative numbers of mutations vs. $t$. Here, $\mu=10^{-5}$ and  $\gamma=0.9$, so that the cumulative numbers match for $t=T$. (d) Log-probability of joining class I and class IV natural WW domains in $T$ steps with the profile (triangles) and RBM (circles) models. Jumps signal the onset of several new mutations, {\em e.g.} 4 in the mean-field {\em free} path at $T=10$ . 
    }
     \label{fig:sketch_MF}
\end{figure}

In the Cont scenario the potential $\pot$ should forbid large jumps along the paths. We thus consider a hard-wall repulsive potential (Fig.~\ref{fig:sketch_MF}(a)),
\begin{equation}
    \Phi_{\text{Cont}} (q) = \frac {\phi(T)}{q-q_c(T)}\ \text{if}\  q_c(T)<q\le 1, \ +\infty \ \text{otherwise.}
\end{equation}
The location of the hard wall, $q_c(T)=1 - \gamma/ T$, allows the path to explore at most $K\equiv T\times N(1-q_c)=\gamma N$ mutations in $T$ steps. Choosing $\gamma\ge D/N$ ($D$ being the Hamming distance between $\mathbf{v}_{start}$ and $\mathbf{v}_{end}$),  is therefore sufficient to interpolate between the two edge sequences, with larger values of $\gamma$ authorizing more flexible paths. The proportionality constant $\phi(T)= 1/T^2$ is set to guarantee the existence of a well defined limit for large $T$.

In the Evo scenario, the potential should emulate Kimura's model of neutral evolution \cite{kimuraneutral}, while the $P_{model}$ factors in Eq.~\ref{eq:path_weight} correspond to selection. Denoting the mutation rate (over a time interval corresponding to one step of the path)  by $\mu$,  the potential is given by \cite{leuthausser1987statistical}
\begin{equation}
    \Phi_\text{Evo} (q) =( 1-q) \ln \left( 1 + \frac{A}{e^{\mu A/(A-1)}-1}\right) \ ,
\end{equation}
where $A=21$ is the number of amino acids plus the gap state; a derivation of $\Phi_\text{Evo}$ can be found in Supplemental Material, Sec.~\ref{SI:neutral}. This potential is linearly decreasing with $q$, see Fig.~\ref{fig:sketch_MF}(a).

Cont and Evo mean-field paths between class-specific WW domains are shown in Fig.~\ref{fig:sketch_MF}(b); both follow similar traces in the specificity plane, in agreement with the paths in Fig.~\ref{fig:WW}(a). However, mutations are homogeneously spread along the Cont path, with $\simeq N\gamma/T$ mutations at each step (Fig.~\ref{fig:sketch_MF}(c-d)). Conversely, the Evo path is highly heterogeneous, with some steps accumulating many mutations and others barely any; see Supplemental Material, Sec.~\ref{SI:consensus} for the list of consensus sequences computed with mean-field theory. Interestingly, most steps along the Evo path I$\to$IV are concentrated in the region characterized by promiscuous sequences binding both ligand classes as mentioned above. The linearity of $\Phi_\text{Evo}$ makes the transition probabilities $\pi$ in ${\cal P}$ in Eq.~\eqref{eq:path_weight} independent of the location of mutations, concentrating intermediate sequences in the region of highest fitness.

\paragraph{Mean-field based estimation of evolutionary distance.}
As an application of our mean-field approach we show how it can be used to estimate evolutionary distances between sequences with complex data-driven models, including epistatic interactions between residues. The probability that sequence $\mathbf{v}_{end}$ be reached after $T$ steps of stochastic mutations with rate $\mu$ starting from $\mathbf{v}_{start}$  is given by
\begin{equation}
    P(\mathbf{v}_{start}\to\mathbf{v}_{end}|T) \sim \exp\left[- N(f^{constrained}_{path} - f^{free}_{path} )\right]\,,
    \label{eq:ev_time}
\end{equation}
where $f^{constrained}_{path}$ is the free energy in Eq.~\ref{eq:free_energy} (with potential $\Phi_{Evo}$) minimized under boundary conditions matching both $\mathbf{v}_{start}$ and $\mathbf{v}_{end}$, while $f^{free}_{path}$ is obtained by releasing the boundary condition at the end extremity of the path. Details on the numerical optimization are given in Supplementary Material, Sec.~\ref{SI:energy_optimisation}. 

This probability can be computed as a function of $T$ to determine the optimal time (evolutionary distance) $T^*$ at which it is maximal. For purely neural evolution,  $f^{free}_{path} = 0$ and the  probability $P(\mathbf{v}_{start}\to\mathbf{v}_{end}|T)$ can be exactly computed;  $T^*$ then coincides with the predictions of Kimura's theory of neutral evolution~\cite{kimuraneutral}, see Supplemental Material, Sec.~\ref{SI:neutral}. $T^*$ can also be easily computed for profile models  \cite{felsenstein2004inferring}, where selection acts independently from site to site, see Fig.~\ref{fig:sketch_MF}(e) for an illustration of WW. Our mean-field theory theory allows us to go well beyond profile models, and to compute the probability $P$ in the presence of epistatic effects in the RBM model inferred from WW sequence data. Figure~\ref{fig:sketch_MF}(e) shows that the evolutionary distance $T^*$ may then substantially differ from its profile counterpart, showing the effectiveness of our mean-field approach to deal with complex sequence models.

\paragraph{Conclusion.} Proteins with known (annotated) functional specificity form a tiny subset of available sequences. Learning accurate, generative class-specific models from these limited data is generally not possible \footnote{Our class-specific RBMs in Fig.~\ref{fig:WW} were used to assess membership to a class, a much easier task than design. Membership to families in PFAM are decided by Hidden Markov Models, which neglect correlations between residues.}. Our path-based approach, inspired by evolutionary dynamics, circumvents this issue and offers an effective way to design proteins interpolating between different functional sub-classes without annotated sequences (apart the anchors of the paths).

In addition, we have introduced a mean-field analysis of paths generated by RBM, characterizing the trajectories of the inputs to the hidden units and of the overlaps between successive sequences. Mean field is a powerful computational scheme in the presence of strong interactions between residues, {\em e.g.} to estimate evolutionary distances. This result opens the way to ancestral reconstruction and to the prediction of phylogenetic trees \cite{felsenstein2004inferring} with data-driven, epistatic  models.

A potentially interesting biological finding in our study of the WW domain is that paths interpolating between classes I and IV  go through a region apparently deprived of natural sequences, albeit corresponding to high RBM likelihood \footnote{We stress that RBM trained on all sequence data, mixing several classes, cannot detect a change of specificity, contrary to RBM models restricted to each class (Fig.~\ref{fig:WW}(f,g)).} and high AlphaFold/ProteinMPNN scores for both ligands (Fig.~\ref{fig:WW}). While experimental investigations are needed to check our finding, these intermediate sequences are putatively unspecialized, and possibly similar to ancestral proteins \cite{promiscuity}\footnote{Additional discussion can be found in Supplemental Material at [url], which includes
Refs.~\cite{jax2018github,MPNN2022,espanel1999single,fischer2012introduction,Jacquin_LP_2016,kimuraneutral,Lau1989Oct,levitt1998unified,Miyazawa1996Mar,Russ2020Jul,tubiana_learning_2019,Tieleman2008Jul,Tubiana2018Nov,tubiana_github,TMscore}.}.

\paragraph{Acknowledgements.}
We are particularly grateful to J. Tubiana for his help on the use of ProteinMPNN. We also thank M. Bisardi, A. Di Gioacchino, A. Murugan, and F. Zamponi for discussions. This work was supported by the ANR-17 RBMPro and ANR-19 Decrypted CE30-0021-01 projects. E.M. is funded by a ICFP Labex fellowship of the Physics Department at ENS.


\typeout{}
\bibliography{ref3,citations}

\end{document}


\maketitle

\section{Path sampling algorithm}
\label{SI:path_alg}

\subsection{Detailed description}
We introduce below the details of the sampling procedure we use to obtain paths connecting two fixed sequences in a landscape described by the probability distribution $P_{model}$. Starting from a path $\{\mathbf{v}_t\}$, we look at intermediate sequences (starting from $t=1$) and propose a mutation with the constraint that the Hamming distance between $\mathbf{v}_{t-1}$ and $\mathbf{v}_{t+1}$ is not greater than 1. We accept this move with a probability fixed to ensure detailed balance. Different cases have to be considered, depending on the Hamming distance $D_H$ between the new attempted sequence and existing ones (note that hereafter we define $\pi({\bf v},{\bf v}')=\exp{[-N \pot(\frac{1}{N} \sum_i \delta_{v_i,v'_i})]}$):
\begin{itemize}
    \item $D_H(\mathbf{v}_{t-1},\mathbf{v}_{t+1}) = 0$. In this case the new sequence $\mathbf{v}'_t$ can have a single mutation at any site, compared with the two adjacent sequences along the path. Hence, we first draw a random site $i$, then we propose a new sequence $\hat{\mathbf{v}}_{t}$ which is equal to  ${\mathbf{v}}_{t-1}$ at all sites but $i$. The new amino-acid for that site $\hat{v}_t^i$ is drawn from the distribution $\propto P^\beta_{model}(\cdot|\mathbf{v}_{t-1}^{\text{\textbackslash} i})$ (here the amino acids are fixed on all the sites different from $i$). Then if the old sequence $\mathbf{v}_t$ already had a mutation with respect to $\mathbf{v}_{t-1}$ at given site $j$, we accept the new mutated sequence $\hat{\mathbf{v}}_t$ (which is equal to $\mathbf{v}_{t-1}$ apart from the amino acid at site $i$)  with a probability
    \begin{align}
        p_{acc}(\mathbf{v}_t&\to \hat{\mathbf{v}}_t) = \nonumber \\ =\min& \left( 1,\frac{\pi(\mathbf{v}_{t-1},\hat{\mathbf{v}}_t)^{2\beta}\sum_zP^\beta_{model}(M_i^z \mathbf{v}_{t-1})}{\pi(\mathbf{v}_{t-1},{\mathbf{v}_t})^{2\beta}\sum_zP^\beta_{model}(M_j^z \mathbf{v}_{t-1})} \right)\,,
    \end{align}
    where $M_i^z$ indicates the mutation $z$ at site $i$. If $\mathbf{v}_t$ or $ \hat{\mathbf{v}}_t$ are equal to $\mathbf{v}_{t-1}$, then the acceptance probability is $p_{acc}(\mathbf{v}_t\to \hat{\mathbf{v}}_t) = \min(1,\pi(\mathbf{v}_{t-1},\hat{\mathbf{v}}_t)^{2\beta}/\pi(\mathbf{v}_{t-1},{\mathbf{v}_t})^{2\beta})$.

    \item $D_H(\mathbf{v}_{t-1},\mathbf{v}_{t+1}) =1 $. In this case the new sequence $\hat{\mathbf{v}}_t$ can have a single mutation only at the site $i$ where $\mathbf{v}_{t-1}$ and $\mathbf{v}_{t+1}$ are different. At that site, we propose a new mutation from the distribution $\propto P^\beta_{model}(\cdot|\mathbf{v}_{t-1}^{\text{\textbackslash} i})$ and accept it with probability $p_{acc}=\exp[-\Lambda \beta(D_H(\hat{\mathbf{v}}_t,\mathbf{v}_{t-1})+D_H(\hat{\mathbf{v}}_t,\mathbf{v}_{t+1})-D_H({\mathbf{v}}_t,\mathbf{v}_{t-1})-D_H({\mathbf{v}}_t,\mathbf{v}_{t+1}))]$.
    
    \item $D_H(\mathbf{v}_{t-1},\mathbf{v}_{t+1}) =2 $. In this case the previous and subsequent sequence present two mutations at site $i$ and $j$. The new sequence  $\hat{\mathbf{v}}_{t}$ can be of two forms: it can have the same mutation of $\mathbf{v}_{t+1}$ (with respect to $\mathbf{v}_{t-1}$) at site $i$ or at site $j$. Hence, we extract one of the two possibilities with a probability weighted accordingly with $P^\beta_{model}(\hat{\mathbf{v}}_t)$.
    
\end{itemize}

\subsection{Proof of detailed balance}
To prove that dynamics given by our algorithm of path sampling converges to the target distribution we have to prove that it respects detailed balance, i.e. the reversibility of each Markov step. We consider the transition from a path $\{\mathbf{v}_t\}$ to a new path that differ only by one sequence $\mathbf{v}_t'$ at time $t$. we write the detailed balance condition as

\begin{align}
    &\mathcal{P}_{path}(\{\mathbf{v}_t\})p_{trans}(\mathbf{v}_t\to \mathbf{v}_t') = \mathcal{P}_{path}(\{\mathbf{v}_t'\})p_{trans}(\mathbf{v}_t'\to \mathbf{v}_t) \nonumber \\
    &\pi(\mathbf{v}_{t-1},\mathbf{v}_t) \pi(\mathbf{v}_t,\mathbf{v}_{t+1}) P_{model}(\mathbf{v}_t) p_{trans}(\mathbf{v}_t\to \mathbf{v}_t') = \pi(\mathbf{v}_{t-1},\mathbf{v}_t')  \pi(\mathbf{v}_t',\mathbf{v}_{t+1}) P_{model}(\mathbf{v}_t') p_{trans}(\mathbf{v}_t'\to \mathbf{v}_t) \nonumber \\
\end{align}

If $D_H(\mathbf{v}_{t-1}, \mathbf{v}_{t+1})=0$, the new sequence can have a mutation at any site $i$ compared to its neighbour $\mathbf{v}_{t-1}$, while $\mathbf{v}_t$ will have the mutation at another site $j$ (note that $i$ and $j$ can be equal. Hence the transition probability in this case will be 

\begin{equation}
   p_{trans}(\mathbf{v}_t\to \mathbf{v}_t') =\frac{1}{N} \frac{P_{model}(\mathbf{v}_t')}{ \sum_{z=1}^Q P_{model}(M_i^z \mathbf{v}_{t-1})}p_{acc}(\mathbf{v}_t\to {\mathbf{v}}_t'), \ p_{trans}(\mathbf{v}_t'\to \mathbf{v}_t) =\frac{1}{N} \frac{P_{model}(\mathbf{v}_t)}{ \sum_{z=1}^Q P_{model}(M_j^z \mathbf{v}_{t-1})} p_{acc}( {\mathbf{v}}_t'\to \mathbf{v}_t).
\end{equation}
By substituting everything in the detailed balance condition we obtain the acceptance probability described in the main paper. This will hold similarly when $D_H(\mathbf{v}_{t-1}, \mathbf{v}_{t+1})=1$ (with $i=j$). For $D_H(\mathbf{v}_{t-1}, \mathbf{v}_{t+1})=2$, the sequences $\mathbf{v}_t$ and $\mathbf{v}_t'$ can either be equal to $\mathbf{v}_{t+1}$ or $\mathbf{v}_{t-1}$, from which the condition presented in the paper descends.

\section{Lattice Proteins}
\label{SI:LP}

To benchmark the performances of this MC procedure to find good transition paths between two sequences, we test it on Lattice Proteins~\cite{Lau1989Oct}, a well known toy-model for protein structure. We consider a protein sequence of 27 amino acids folding into a 3D structure specified as a self-avoiding path over a 3x3x3 lattice where each amino acid occupies one node. The probability of a sequence $\mathbf{v}$ to fold into a specific structure $\mathcal{S}$ is given by the interaction energies between amino acids in contact in the structure (i.e. those who occupy neighbouring nodes of the lattice, but are not adjacent in the protein sequence). In particular, the total energy of a sequence with respect to a given structure is given by
\begin{equation}
    \mathcal{E}_{LP}(\mathbf{v}|\mathcal{S}) = \sum_{i<j} c_{ij}^\mathcal{S}E_{MJ}(v_i,v_j)\,
\end{equation}
where $c^\mathcal{S}$ is the contact map ($c_{ij}^\mathcal{S}=1$ if sites are in contact and $0$ otherwise), while the pairwise energy $E_{MJ} ( v_i , v_j )$ represents the amino-acid physico-chemical interactions given by the the Miyazawa-Jernigan knowledge-based potential~\cite{Miyazawa1996Mar}. The probability to fold into a specific structure is written as
\begin{equation}
    p_{nat}(\mathcal{S}|\mathbf{v}) = \frac{e^{-\mathcal{E}_{LP}(\mathbf{v}|\mathcal{S})}}{\sum_{\mathcal{S}'}e^{-\mathcal{E}_{LP}(\mathbf{v}|\mathcal{S'})}}\, ,
\end{equation}
where the sum is over the entire set of self-avoiding path in the cubic lattice.\\
The function $p_{nat}$ represents a suitable landscape that maps each sequence to a score measuring the quality of its folding. To study in more detail this landscape, we will consider an alignment of sequences folding into a specific structure (that we will call $\mathcal{S}$) sampled from a low temperature MC sampling using $-\beta\log p_{nat}(\cdot|\mathcal{S})$ (with $\beta=10^3$) as effective energy ~\cite{Jacquin_LP_2016}.

\section{Restricted Boltzmann Machines and training parameters}
\label{SI:RBMtrain}
To study the problem of transition paths we first need a model to infer a landscape from our sequence data set. At this scope, we are going to use Restricted Boltzmann Machines, an unsupervised energy-based model able to learn representations of the data in a two-layer bipartite graph~\cite{Fischer2012Sep}. The first "visible" layer represents the protein sequence $\mathbf{v}=\{v_1,...,v_N\}$ where each unit takes one out of 21 possible states (20 amino acids + 1 alignment gap). The second is the "hidden" layer which displays the real-valued representations $\mathbf{h}=\{h_1,...,h_M\}$. The joint probability distribution for $\mathbf{v}$ and $\mathbf h$ is
\begin{equation}
    P_{RBM}(\mathbf{v},\mathbf{h}) \propto \exp\left[ \sum_{i=1}^Ng_i(v_{i}) + \sum_{i,\mu}w_{i\mu}(v_{i})h_{\mu}-\sum_\mu\mathcal{U}(h_{\mu})\right]
    \label{joint_dist_rbm}
\end{equation}
up to a normalization constant. visible and hidden units are coupled through the matrices $w_{i\mu}$ and the value of each unit is biased by the local fields $g_i$ and $\mathcal{U}_\mu$.
In \cite{tubiana_learning_2019} it has been shown that this model is able to recover statistically relevant sequence motifs playing crucial roles in the structure and functionality of different protein families.
Following their approach, we choose $\mathcal{U}_\mu$ to be double Rectified Linear Unit (dReLU) potentials of the form
\begin{equation}
    \mathcal{U}_\mu(h) = \frac{1}{2}\gamma_{\mu,+}h_+^2 + \frac{1}{2}\gamma_{\mu,-}h_-^2 + \theta_{\mu,+}h_+ + \theta_{\mu,-}h_-\, ,\ \text{where } h_+ = \max(h,0),\ h_- = \min(h,0),
    \label{dRelu_rbm}
\end{equation}
where we have defined the hyper-parameters $\gamma_{\mu,\pm}$, $\theta_{\mu,\pm}$.\\
At this point, we need a learning procedure to infer the hyper-parameters that best fit our data. We decided to use a Persistent Contrastive Divergence algorithm~\cite{Tieleman2008Jul} which has been shown to be sufficiently good and robust under cautious choice of the regularization hyper-parameters~\cite{Tubiana2018Nov}. The code and the data used to train our RBMs for Lattice Proteins and WW domains can be found in~\cite{tubiana_github}. The hyper-parameters used for learning are the following:

\begin{itemize}
    \item For WW (N=31):
    \begin{itemize}
        \item M = 50
        \item Batch size = 100
        \item Number of epochs = 500
        \item Learning rate =  $5 \times 10^{-3}$ (which has a decay rate of 0.5 after 50\% of iterations) 
        \item $L_1b$ regularization = 0.25
        \item Number of MC step between each update = 10
    \end{itemize}
      \item For Lattice Protein ($N$=27):
    \begin{itemize}
        \item $M = 100$
        \item Batch size = 100
        \item Number of epochs = 100
        \item Learning rate = $5 \times 10^{-3}$ (which has a decay rate of 0.5 after 50\% of iterations) 
        \item $L_1b$ regularization = 0.025
        \item Number of MC step between each update = 5
    \end{itemize}
\end{itemize}
For the local RBMs trained respectively on the three specificity classes of WW sequences predicted by the original RBM, we use $M=30$ and keep all the other hyper-parameters unchanged.

\section{Statistics of paths sampled for LP}
\label{SI:LP_RBM_stats}
\begin{figure}
    \centering
    \includegraphics[width=0.7\linewidth]{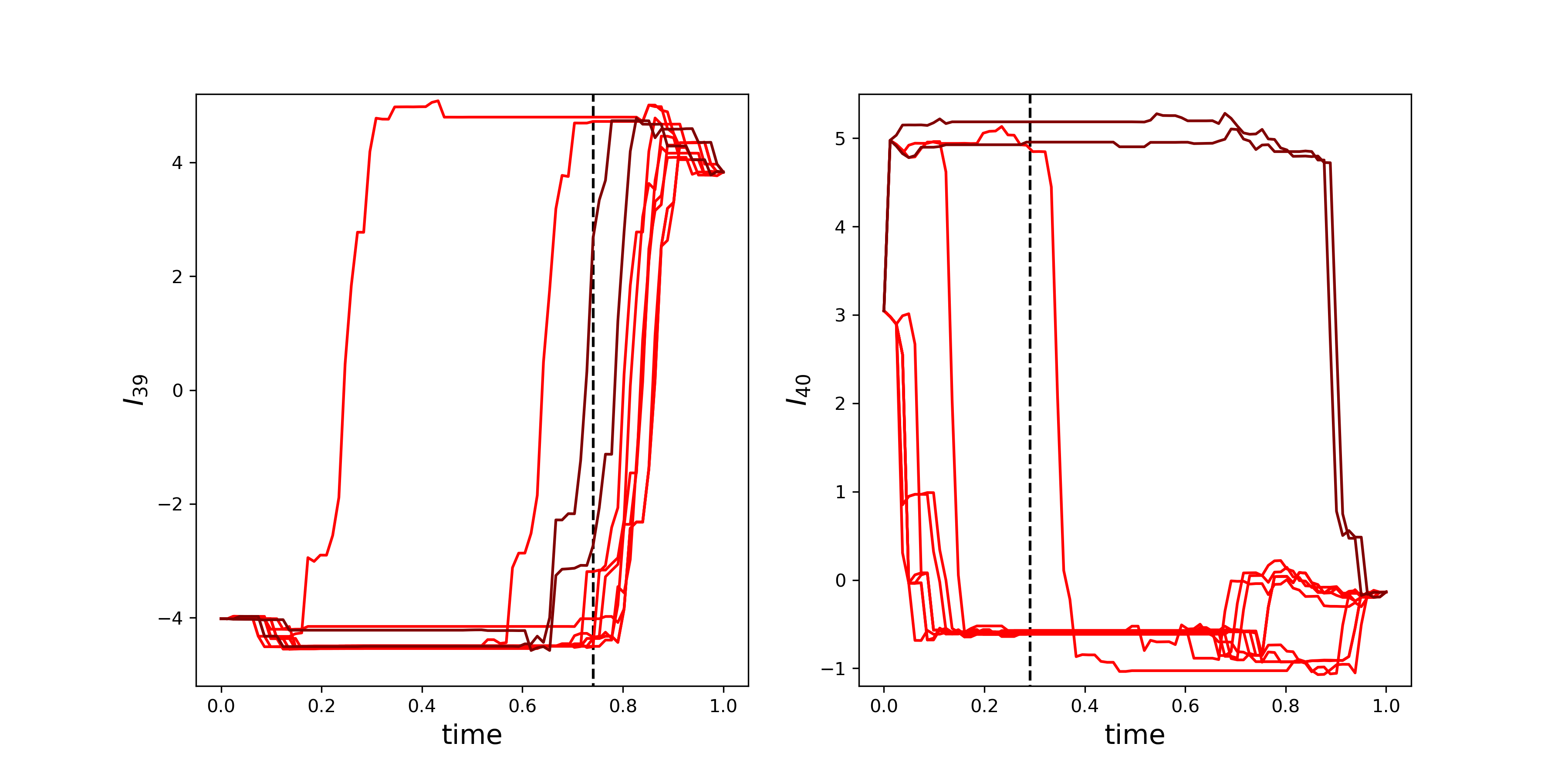}
    \caption{Plot of some relevant inputs as function of time sampled with our Monte-Carlo procedure ($\Lambda=2$ and target $\beta=3$). The colors respect those presented in Figure 1 in the main paper. Black lines correspond to the average time at which the input switch value (>0 for $I_4$ and <2 for $I_{14}$).}
    \label{fig:stat}
\end{figure}

\begin{figure}
    \centering
    \includegraphics[width=0.6\linewidth]{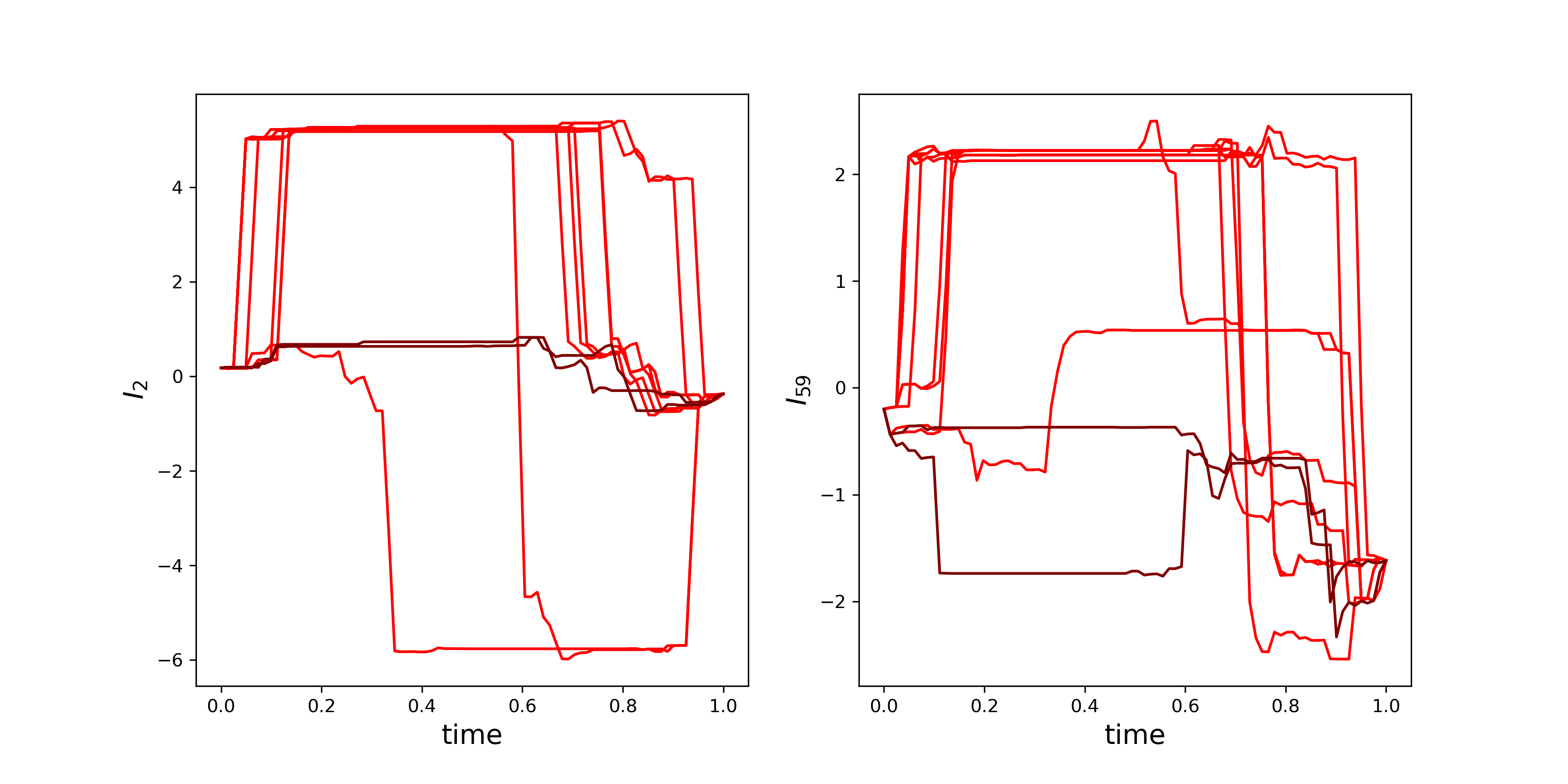}
    \hspace{0.1cm}
    \includegraphics[width=0.35\linewidth]{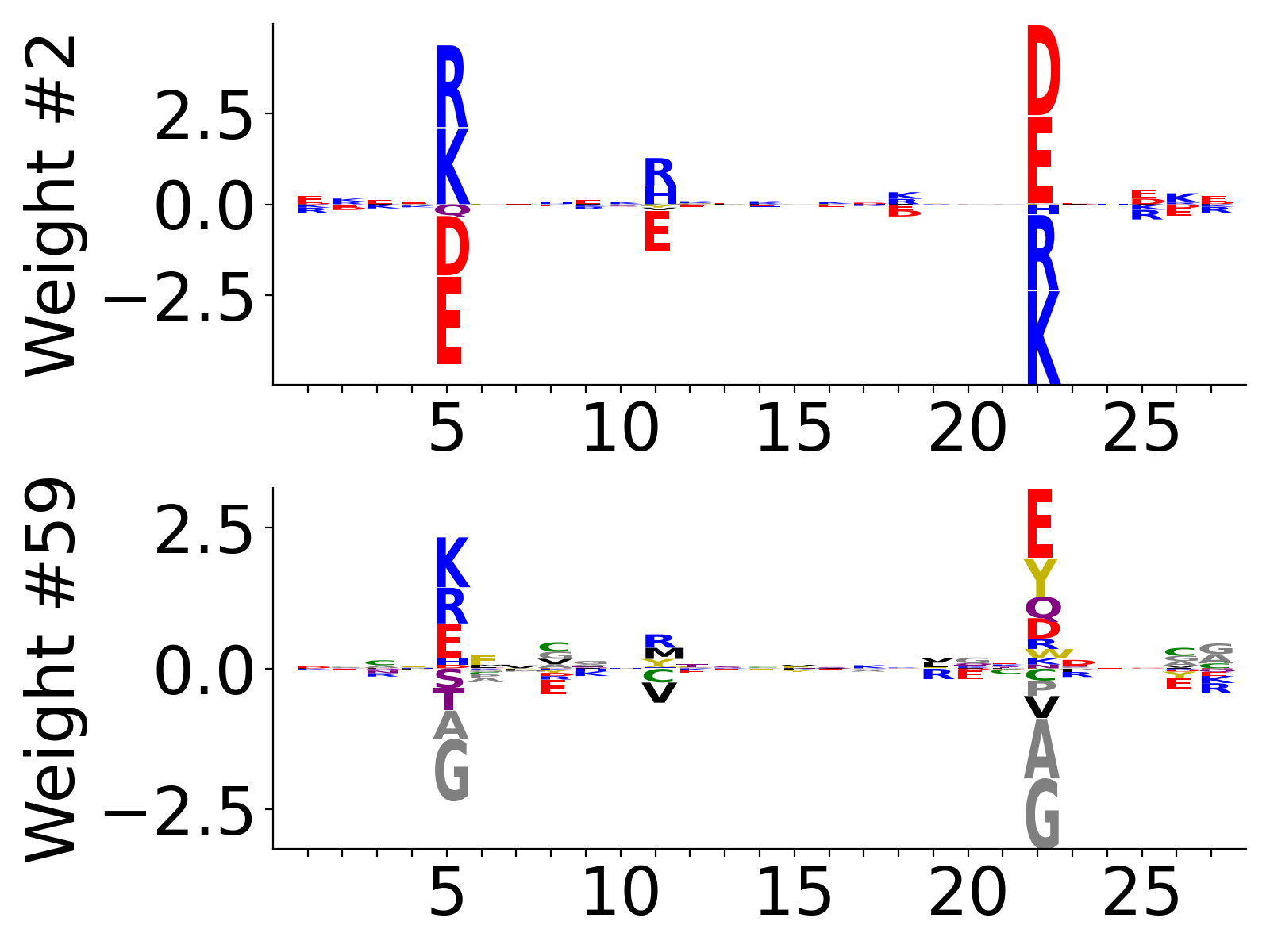}
    \caption{Plot of some relevant inputs (and their respective weights logos) as function of time sampled with our Monte-Carlo procedure ($\Lambda=2$ and target $\beta=3$) exploiting relevant electrostatic interactions not exploited in the edge sequences {$\bf v_{start}$} and {$\bf v_{end}$}. The colors respect those presented in Figure 1 in the main paper.}
    \label{fig:stat2}
\end{figure}

Hereafter we report some statistics over the paths sampled in the Lattice Protein model. We observe in Figure~\ref{fig:stat} that many paths prefer to activate input 40, exploiting Cys-Cys interactions (see logos of the weights in the Appendix). Furthermore, statistically this input is more likely to be reduced before activating input 4.\\
Analysing the paths along each inputs we notice that those solutions maintain high scores in terms of $p_{nat}$ by exploiting the interactions between amino acids at sites 5,6,11 and 22 (see Figure 1 in the main paper). These paths are divided in two classes corresponding to the chemical nature of the interaction used to bind these amino-acids. One cluster (shown in maroon in Figure 1 of the main paper and in Figure~\ref{fig:stat}) uses Cys-Cys bridges to establish this interactions. Conversely the most populated cluster (shown in red in the same figures) exploits electrostatic interactions that are not initially present in the target sequences. Some of these interactions are shown in Figure~\ref{fig:stat2}.

\begin{figure}
    \centering
    \includegraphics[width=0.5\linewidth]{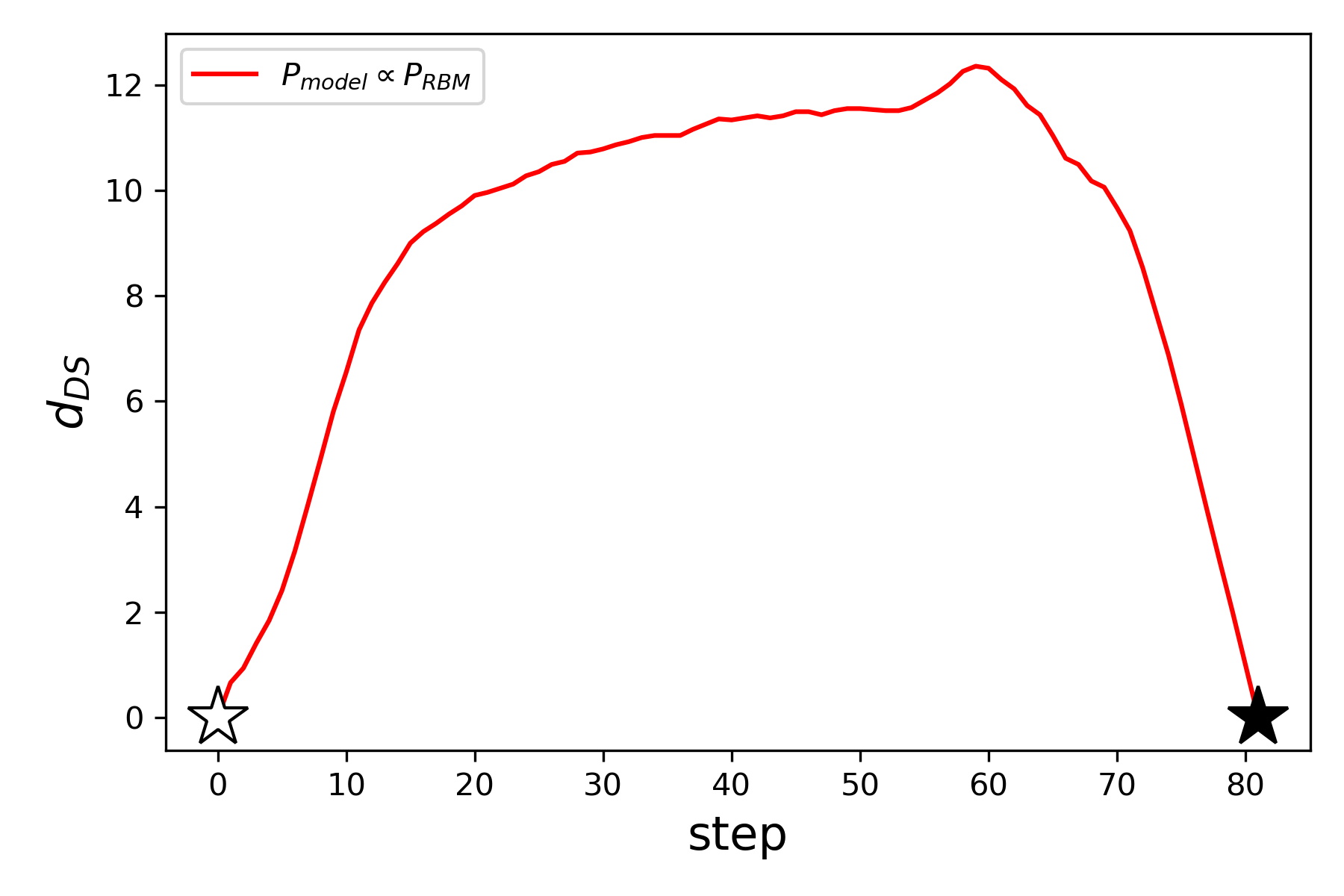}
    \caption{Average value of the distance from direct space (defined as $d_{DS}({\bf v}) = \sum_i (1-\delta_{v^{start}_i,v_i})(1-\delta_{v^{end}_i,v_i})$) as a function of the step along the path connecting the two modes in the LP landscape. Black and White stars refer to the same sequences as in Figure 1 of the main text.}
    \label{fig:dist_from_direct}
\end{figure}

To deeper characterize the behaviour of these solutions, we also plot in Figure \ref{fig:dist_from_direct} the average distance from the two edge sequences (also called distance from direct space) as a function of time obtained from the paths sampled using  $P_{model}\propto P_{RBM}$, the likelihood from the trained RBM model.

\section{Additional information about mutational paths of the WW domain}

\subsection{Lists of the tested sequences}
\label{SI:WW_list}
Here we present the reference sequences sampled with the MC algorithm and tested using AlphaFold (note that the first sequence of each list represent the YAP1 wild-type sequence from~\cite{espanel1999single}, while the last is the natural wild-type specific for each class) together with the predicted specificity using local RBMs:
\begin{itemize}
    \item From YAP1 to wild-type protein in class II:
    \begin{itemize}
        \item \texttt{LPAGWEMAKTSS-GQRYFLNHIDQTTTWQDP} 
        \item \texttt{LPAGWEMAKTS\hl{D}-G\hl{E}RYF\hl{I}NH\hl{N}\hl{T}\hl{K}TTTWQDP} predicted I
        \item \texttt{LP\hl{P}GWE\hl{E}A\hl{R}T\hl{P}\hl{D}-GR\hl{V}YF\hl{I}NH\hl{N}\hl{T}\hl{K}TTTWQDP} predicted I
        \item \texttt{LP\hl{P}GWE\hl{E}A\hl{R}\hl{A}\hl{P}\hl{D}-GRTY\hl{Y}YNH\hl{N}\hl{T}\hl{K}TTTW\hl{E}\hl{K}P} predicted II/III
        \item \texttt{LP\hl{P}GW\hl{T}\hl{E}\hl{H}K\hl{A}\hl{P}\hl{D}-GRTY\hl{Y}YNH\hl{N}\hl{T}\hl{K}T\hl{S}TW\hl{E}\hl{K}P} predicted II/III
        \item \texttt{LPSGW\hl{T}\hl{E}\hl{H}K\hl{A}\hl{P}\hl{D}-GRTY\hl{Y}YN\hl{T}E\hl{T}\hl{K}\hl{Q}\hl{S}TW\hl{E}\hl{K}P} predicted II/III
        \item \texttt{\hl{A}\hl{K}S\hl{M}W\hl{T}\hl{E}\hl{H}KSP\hl{D}-GRTY\hl{Y}YN\hl{T}E\hl{T}\hl{K}\hl{Q}\hl{S}TW\hl{E}\hl{K}P}
    \end{itemize}
    \item From YAP1 to wild-type protein in  class IV:
    \begin{itemize}
        \item \texttt{LPAGWEMAKTSS-GQRYFLNHIDQTTTWQDP}
        \item \texttt{LPAGWEM\hl{R}\hl{R}T\hl{P}S-G\hl{R}\hl{V}YF\hl{V}NHI\hl{T}\hl{R}TT\hl{Q}W\hl{E}DP} predicted I
        \item \texttt{LP\hl{P}GWE\hl{E}\hl{R}\hl{R}\hl{D}\hl{P}S-G\hl{R}\hl{V}Y\hl{Y}\hl{V}NHI\hl{T}\hl{R}TT\hl{Q}W\hl{E}\hl{R}P} predicted I
        \item \texttt{LP\hl{P}GWE\hl{E}\hl{R}VSRS-G\hl{R}\hl{V}Y\hl{Y}\hl{V}NHI\hl{T}\hl{R}TT\hl{Q}W\hl{E}\hl{R}P} predicted I
        \item \texttt{LP\hl{P}GWEK\hl{R}\hl{M}SRS-G\hl{R}\hl{V}Y\hl{Y}\hl{V}NHI\hl{T}\hl{R}TT\hl{Q}W\hl{E}\hl{R}P} predicted I
        \item \texttt{LP\hl{P}GWEK\hl{R}\hl{M}SRSSG\hl{R}\hl{V}Y\hl{Y}\hl{V}NHI\hl{T}\hl{R}\hl{A}\hl{S}\hl{Q}W\hl{E}\hl{R}P} predicted IV
        \item \texttt{LP\hl{P}GWEK\hl{R}\hl{M}SRSSG\hl{R}\hl{V}Y\hl{Y}FNHI\hl{T}N\hl{A}\hl{S}\hl{Q}W\hl{E}\hl{R}P}
    \end{itemize}
    \item From YAP1 to wild-type protein in  class I:
    \begin{itemize}
        \item \texttt{LPAGWEMAKTSS-GQRYFLNHIDQTTTWQDP}
        \item \texttt{LPAGWEMAKTS\hl{E}-GQRYF\hl{I}NH\hl{N}TQTTTWQDP}
        \item \texttt{LP\hl{P}GWEMA\hl{Y}T\hl{P}\hl{E}-G\hl{E}RYF\hl{I}NH\hl{N}\hl{T}\hl{K}TTTW\hl{L}DP}
        \item \texttt{LP\hl{P}GWEM\hl{G}ITRG-GR\hl{V}\hl{F}F\hl{I}NHE\hl{T}\hl{K}\hl{S}TTW\hl{L}DP}
        \item \texttt{LPR\hl{S}W\hl{T}\hl{Y}\hl{G}ITRG-GR\hl{V}\hl{F}F\hl{I}NHEA\hl{K}\hl{S}TTW\hl{L}\hl{H}P}
        \item \texttt{LPR\hl{S}W\hl{T}\hl{Y}\hl{G}ITRG-GR\hl{V}\hl{F}F\hl{I}NEEA\hl{K}\hl{S}TTW\hl{L}\hl{H}P}
    \end{itemize}
\end{itemize}

\subsection{Additional paths associated to I$\to$IV transition}
\label{SI:ItoIVadditionalstatistics}
\begin{figure}
    \centering
    \includegraphics[width=\linewidth]{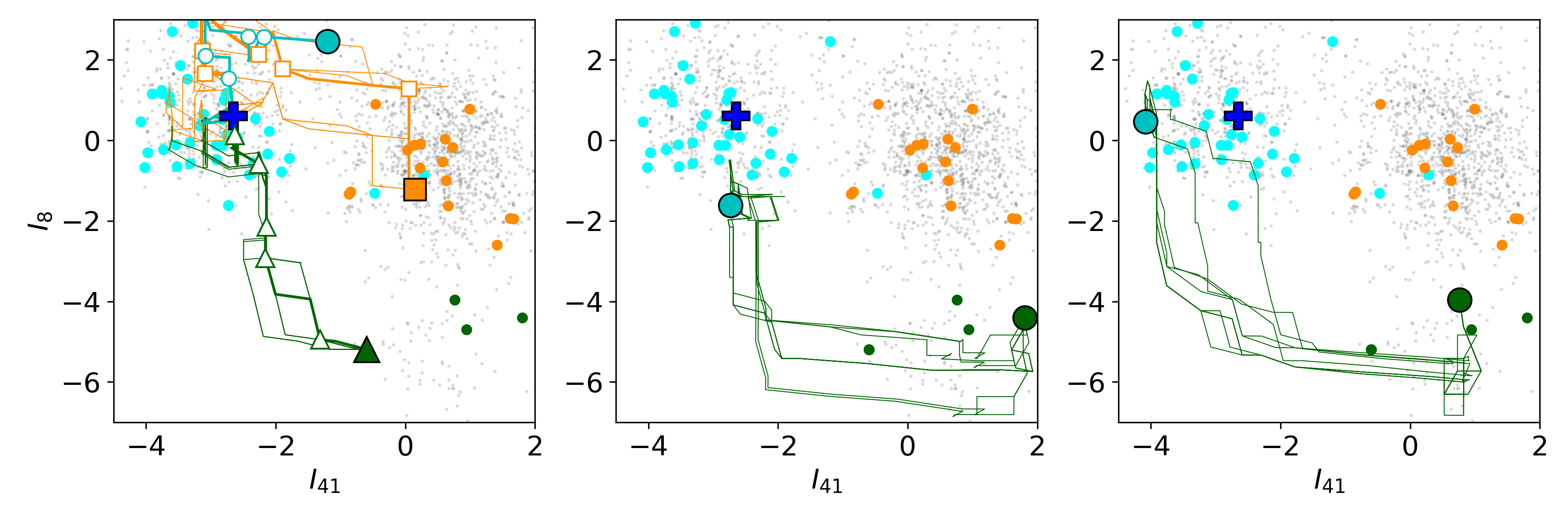}
    \caption{(a) Same as Figure 2 in the main paper. 10 new samples of paths connecting specificity class I to class IV (PDB IDs: 2LTW and 1I8G, respectively) and class I to class II/III are shown. (b)-(c) 10 paths joining two pairs of wild-type sequences in classes I and IV obtained with the same sampling algorithm as in Figure 2 in the main text.}
    \label{fig:SI_statistics_path}
\end{figure}

We show in Figure~\ref{fig:SI_statistics_path} different paths sampled for different couples of class I and class IV wild-types using the same simulation parameters as in the main paper. This figure shows that all the paths cross a region of the input space where natural sequences are lacking.

\subsection{Details on the TM-score}
\label{SI:WW_TM_score}

\begin{figure}[H]
    \centering
    \includegraphics[width=0.9\linewidth]{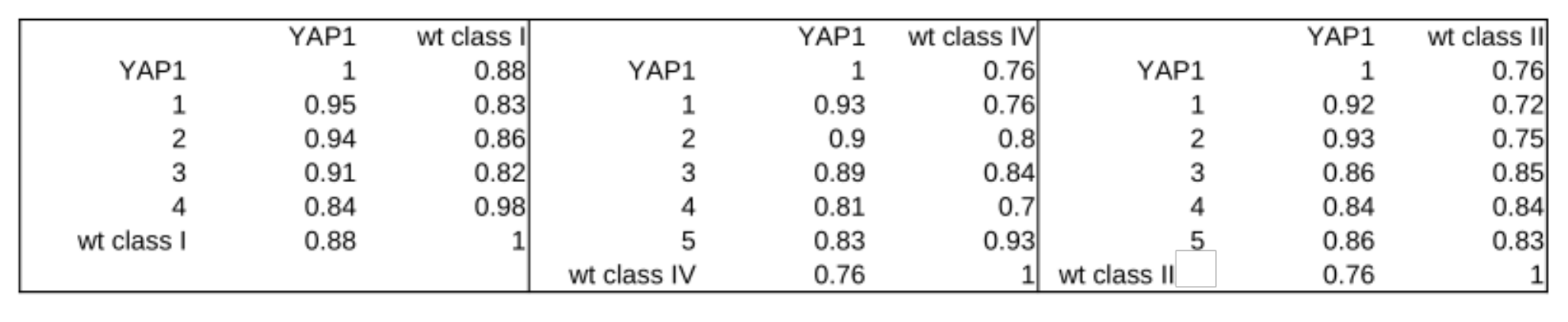}
    \caption{Table of the TM-scores measured between the structures inferred from AlphaFold.}
    \label{fig:TMscore}
\end{figure}

To compare the inferred structures of the sampled sequences along the path with those of the target natural sequences we used the Template Modelling (TM) score developed in~\cite{TMscore} and represents a variation of the Levitt–Gerstein (LG) score~\cite{levitt1998unified}. Compared to other similarity score (like root-mean-square deviation (RMSD)) it gives a more accurate measure since it relies more on the global similarity of the full sequence rather than the local similarities.\\  
Practically, we consider a target sequence of length  $L_{\text{target}}$ and a template one whose structure has to be compared with. First, we align the two sequences and we take the $L_{\text{common}}$ pairs of residues that commonly appear aligned. Then the score is computed as
\begin{equation}
    \text{TM-score} = \max_{\{d_i\}}\left[ \frac{1}{L_\text{target}}\sum_{i=1}^{L_\text{common}}\frac{1}{1+\left( \frac{d_i}{d_0(L_\text{target})}\right)^2}\right]\,,
\end{equation}
where $d_{i}$ is the distance between the $i$th pair of residues between the template and the target structures after alignment, and  $d_{0}(L_{\text{target}})=1.24(L_{\text{target}}-15)^{1/3}-1.8$ is a distance scale that normalizes distances. This formula gives a score between 0 and 1. if TM-score$<0.2$, the two sequences are totally uncorrelated, while they can be considered to have the same structure if TM-score$>0.5$. In Figure \ref{fig:TMscore} we show the tables of the TM-scores associated with the sequences tested above.

\subsection{ProteinMPNN score}
\label{SI:MPNN}
The ProteinMPNN model published in~\cite{MPNN2022} takes a reference backbone structure (uploaded as a .pdb file) and gives as output a log-probability function over protein sequences $\log P_{MPNN}(\mathbf{v})$ measuring the affinity of the sequence $\mathbf{v}$ to fold into a specific structure and/or complex. PDB IDs: 2LTW for class I wild-type , 1YWI for class II wild-type, 1I8G for class IV wild-type.  

In order to compute the likelihood of a sequence given a reference backbone, the sequence has to have the same length of the wild-type from which the reference backbone was taken. Unfortunately, the transition from class I to class IV requires an insertion along the path (see section \ref{SI:WW_list}). So, we decided to remove those insertion to test the sequences against the class I reference backbone structure (PDB ID: 2LTW) and to substitute the gap with Serine (S) to test them against class IV reference structure (PDB ID: 1I8G).

\begin{figure}
    \centering
    \includegraphics[width=\linewidth]{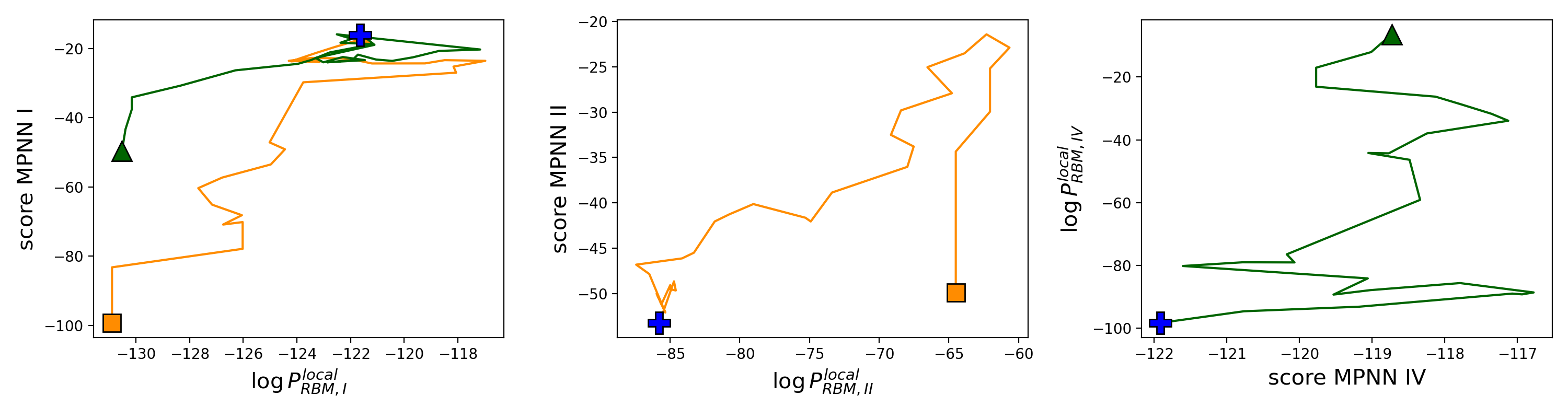}
    \caption{Comparison of the MPNN score and local RBM likelihoods along the paths shown in Figure 2 of the main paper.}
    \label{fig:SI_MPNN_RBM_compared}
\end{figure}

We compare the value of the MPNN score along the paths with the likelihood measured by the RBM locally trained on the different specificity classes in Figure~\ref{fig:SI_MPNN_RBM_compared}. The plot shows a slight correlation between the two scores.

\section{Derivation of mean-field equations for path sampling with a RBM model}
\label{SI:RBM_MF_eqs}
To exploit the nature of the RBM as a mean-field model we rewrite the probability distribution of the model as 
\begin{align}
    P_{RBM}(\textbf{v}) &= \frac{1}{Z_{RBM}}\int \prod_\mu \text{d}h_\mu\exp\left( \sum_ig_i(v_i) + \sum_{i,\mu}w_{i\mu}(v_i) h_\mu - \sum_\mu \mathcal{U}_\mu(h_\mu) \right)\\ &= \frac{1}{Z_{RBM}}\exp\left( \sum_i g_i(v_i) + N \sum_\mu\Gamma_\mu \big( \frac 1N I_\mu \big) \right)\, ,
\end{align}
where $\gamma$ is defined in the main text.

By introducing the order parameters $m_t^\mu = I_{\mu,t}/N$, $q_t = \frac{1}{N}\sum_i\delta_{v_{i,t},v_{i,t+1}}$ as well as the overlap potential $\Phi$, we can write the probability for a path in the order parameter space as (in the large $N$ limit)
\begin{align}
    P_{path}(\{\textbf{m}_t,q_t\}_{t=1}^{T};\beta) &\propto \sum_{\{\mathbf{v}_t\}}\prod_{t,\mu} \delta\left(\frac{1}{N}\sum_i w_{i\mu}(v_{i,t})- m_t^\mu\right) \prod_{t}\delta\left(\frac{1}{N}\sum_i \delta_{v_{i,t},v_{i,t+1}}- q_t\right) P^\beta_{RBM}(\mathbf{v}_t)\pi^\beta(\mathbf{v}_t,\mathbf{v}_{t+1}) \\ &= \exp\left[\beta N \left( \sum_{t,\mu}{\Gamma( m_t^\mu)} - \sum_t\pot(q_t)\right)\right ] \times \\ \times \int \left( \prod_{\mu,t} \text{d}\hat{m}_t^\mu \prod_t\text{d}\hat{q}_t\right )&\sum_{\{\mathbf{v}_t\}}\exp\left[\beta \sum_{i,t} g_i(v_{i,t}) + N\sum_{t,\mu} \hat{m}_t^\mu \left(\frac{1}{N}\sum_i w_{i\mu}(v_{i,t})- m_t^\mu \right) + N \sum_t \hat{q}_t \left(\frac{1}{N}\sum_i \delta_{v_{i,t},v_{i,t+1}}- q_t\right) \right] \\ &\approx \exp\left[\beta N \left( \sum_{t,\mu}{{\Gamma}( m_t^\mu)} - \sum_t\pot(q_t)\right) +N\mathcal{S}(\{\textbf{m}_t,q_t\}_{t=1}^{T}) \right] = \exp\left(- N\beta f_{path}(\{\textbf{m}_t,q_t\}) \right)\, ,
\end{align}

where

\begin{equation}
    \mathcal{S}(\{\textbf{m}_t,q_t\}_{t=1}^{T}) = \min_{\{\hat{\textbf{m}}_t,\hat{q}_t\}} \frac{1}{N} \sum_i \log {Z}_i(\{\hat{\textbf{m}}_t,\hat{q}_t\}) - \sum_{t,\mu} m_t^\mu\hat{m}_t^\mu - \sum_t q_t\hat{q}_t
\end{equation}

and

\begin{equation}
    {Z}_i(\{\hat{\textbf{m}}_t,\hat{q}_t\}) = \sum_{v_1,..,v_t,...,v_{T}} \exp \left[ \beta \sum_t g_i(v_t) + \sum_{t,\mu} \hat{m}_t^\mu w_{i\mu}(v_t) + \sum_t \hat{q}_t\delta_{v_t,v_{t+1}}\right]\,.
\end{equation}
Under minimization we find the result shown in the main paper. To obtain numerically the set of magnetizations and overlap that minimize the free energy, we note that the saddle point equation for $f_{path}$ leads to the following self-consistent equation:
\begin{align}
    m_t^\mu &= \frac{1}{N}\sum_i \frac{1}{{Z}_i} \sum_{v_1,..,v_t,...,v_{T}} w_{i\mu}(v_t)\exp \left[ \beta \sum_t g_i(v_t) + \sum_{t,\mu} \hat{m}_t^\mu w_{i\mu}(v_t) + \sum_t \hat{q}_t\delta_{v_t,v_{t+1}}\right]\\
    q_t &= \frac{1}{N}\sum_i \frac{1}{{Z}_i} \sum_{v_1,..,v_t,...,v_{T}} \delta_{v_t,v_{t+1}}\exp \left[\beta \sum_t g_i(v_t) + \sum_{t,\mu} \hat{m}_t^\mu w_{i\mu}(v_t) + \sum_t \hat{q}_t\delta_{v_t,v_{t+1}}\right]\,,
\end{align}
where $\hat{q}_t = - \beta \pot'(q_t)$ and $\hat{m}_t^\mu = \beta {\Gamma}_\mu'(m_t^\mu)$. We solve this set of equations using gradient descent. To compute the LHS we first compute the partition functions ${Z}_i$ using the transfer matrix method and then we take their gradient using automatic differentiation technique built in the Python library JAX~\cite{jax2018github}.\\

\subsection{Consensus sequence from MF solutions and the case for WW domain}
\label{SI:consensus}
To obtain the average distance from the direct space, we need to compute at each time at each site the probability of a specific state $a=1,...,A$. This can be computed as

\begin{align}
     &f_{i,t}(a|\{\textbf{m}_t,q_t\}) = \parder{\beta g_{it}(a)} \log Z_{path} =  \parder{\beta g_{it}(a)} \sum_\pathV \exp \left[ \sum_{i,t,a} \beta g_{it}(a)\delta_{a,v_{it}} + N\beta\sum_{\mu,t}\Gamma(m_\mu^t) - N\beta \sum_t \pot(q_t) \right] = \parder{\beta g_{it}(a)} \log Z_i\,,
\end{align}
where we write $g_{it}(a) = g_i(a)$ for any $t$. Once $f_{i,t}$ is computed, we obtain the consensus sequence ${\bf v}_t = \{ v_{i,t} = \text{argmax}_a f_{i,t} \}$. The consensus sequences for the MF path shown in Figure (3) of the main paper are:

\begin{itemize}
\item I$\to$IV Cont scenario:
      \begin{itemize}
      \item \texttt{LPAGWEMAKTSS-GQRYFLNHITRTTTWQDP}
      \item \texttt{LPAGWEMRKTSS-GQVYFLNHITRTTTWEDP}
      \item \texttt{LPPGWEMRKTSS-GRVYFLNHITRTTQWEDP}
      \item \texttt{LPPGWEMRKSRS-GRVYFLNHITRTTQWEDP}
      \item \texttt{LPPGWEKRKSRS-GRVYYLNHITRTTQWERP}
      \item \texttt{LPPGWEKRMSRS-GRVYYLNHITRTTQWERP}
      \item \texttt{LPPGWEKRMSRS-GRVYYFNHITRASQWERP} 
      \end{itemize}
\item I$\to$IV Evo scenario:
 \begin{itemize}
      \item \texttt{LPPGWEKRKSRS-GRVYYLNHITKTTQWERP}
      \item \texttt{LPPGWEKRKSRS-GRVYYLNHITKTTQWERP}
      \item \texttt{LPPGWEKRKSRS-GRVYYLNHITKTTQWERP}
      \item \texttt{LPPGWEKRKSRS-GRVYYLNHITKTTQWERP}
      \item \texttt{LPPGWEKRKSRS-GRVYYLNHITKTTQWERP}
      \item \texttt{LPPGWEKRMSRS-GRVYYLNHITKTTQWERP}
      \item \texttt{LPPGWEKRMSRS-GRVYYLNHITKTTQWERP} 
      \end{itemize}
\item I$\to$II Cont scenario:
 \begin{itemize}
      \item \texttt{LPAGWEMAKTSD-GQRYFLNHITQTTTWQDP}
      \item \texttt{LPAGWEMAKTPD-GQRYFLNHITKTTTWEDP}
      \item \texttt{LPAGWEEAKTPD-GRTYFYNHITKTTTWEDP}
      \item \texttt{LPAGWEEAKTPD-GRTYYYNHITKTTTWEKP}
      \item \texttt{LPAGWTEHKTPD-GRTYYYNHITKTTTWEKP}
      \item \texttt{LPSGWTEHKTPD-GRTYYYNTITKQSTWEKP}
      \item \texttt{LPSGWTEHKSPD-GRTYYYNTETKQSTWEKP} 
      \end{itemize}
\item I$\to$II Evo scenario:
 \begin{itemize}
      \item \texttt{LPAGWEEAKTPD-GRRYFLNHITKTTTWEDP}
      \item \texttt{LPAGWEEAKTPD-GRTYYYNHITKTTTWEDP}
      \item \texttt{LPAGWEEAKTPD-GRTYYYNHITKTTTWEKP}
      \item \texttt{LPAGWEEAKTPD-GRTYYYNHITKTTTWEKP}
      \item \texttt{LPSGWTEHKTPD-GRTYYYNTITKQSTWEKP}
      \item \texttt{LPSGWTEHKTPD-GRTYYYNTETKQSTWEKP}
      \item \texttt{LPSGWTEHKSPD-GRTYYYNTETKQSTWEKP} 
      \end{itemize}
\end{itemize}

\subsection{Free energy optimisation in mean-field theory}
\label{SI:energy_optimisation}
In order to obtain the minimum of the free energy in the mean-field approximation we first modify the energetic term of the RBM model by multiplying it with temperature factor $\beta_0$ (the interaction term $\sum_t \pot(q_t)$ stays untouched). Starting from $\beta_0=0$, we minimize the free-energy using gradient descent. Here the landscape is convex and the gradient descent finds the global minima. Using this solution as a new initial configuration, we re-use gradient descent but after having increased $\beta_0$ by a small step $\delta \beta_0$. In such a way we can follow the global minimum (under the hypothesis that the system does not encounter zeroth-order phase transitions). Then we repeat this procedure until we reach $\beta_0 = \beta$.

\section{Interpretation of the different parameters}
\label{SI:parameter_interp}
We hereafter recall the meaning of the different parameters we introduced in the main text and the rationales for their values:
\begin{itemize}
    \item[$\Lambda$] is a penalty term we introduce in the sampling algorithm to control the number of mutations accumulated along the path. Very high values of this parameter minimizes the total number of mutations along the path, which will be equal to the Hamming distance $D$ between the target sequences. Hence, reverse mutations become statistically unlikely, limiting the exploration of the sequence space  and  resulting in intermediate sequences with generally lower scores according to $P_{model}$. Hence, we chose $\Lambda \sim 0.1-1$ In order to guarantee a deeper exploration of the sequence space, while disfavoring reverse transitions and obtaining smoother paths.
    \item[$\gamma$], similarly to $\Lambda$,  controls the maximum number of mutations available at each step along the path in the mean-field model (Cont scenario). In the main paper (see Figure 3), $\gamma$ has been chosen to match the number of total mutations at the end of the path between the Evo and Cont scenario (see Figures 3c and 3d).
    \item[$\beta$] represents the inverse equilibrium temperature at we which we set the system composed by the transition path which can be seen as a polymer -- with the two ends are fixed -- evolving at equilibrium through an energy landscape described in Eq. 1 on the main paper. For the sampling algorithm we used $\beta=3$ (see Figures 1 and 2 in the main manuscript). Taking such a low temperature is necessary if we aim to sample sequences that are biologically functional. A similar observation was made in~\cite{Russ2020Jul}, where novel functional sequences of chorismate mutase (CM) were generated from a Potts model trained on natural homologous sequences. For the mean-field theory, we decided to set $\beta=1$ since we are assuming that the inferred model $P_{model}$ is a good proxy of the real fitness function driving the evolution.
    \item[$\mu$] is the mutation rate of a genome evolving in a fitness landscape defined by $P_{model}$ in the Evo scenario, measured in [mutations/ genome length/ cellular division]. The value of the mutation rate $\mu$ in the main paper has been chosen to be consistent with typical mutation rates of viruses (in the simulations we chose $\mu=10^{-5}$ mutations/genome length/cellular division). We note that for the mean-field model the parameter that really matters is $\mu\times T$ compared with the Hamming distance $D$ between the target sequences.
\end{itemize}

\section{Neutral theory of evolution}
\label{SI:neutral}
Let's consider a sequence (with $A$ number of states per site) evolving under mutations only. Given a site $i$ along the sequence the probability of that site to be in a given state $a$ at time $t$, $x^i_a(t)$, evolves through time under the following equation:
\begin{equation}
    \frac{d}{dt}x^i_a(t) = - \mu  x^i_a(t) + \frac{\mu}{A}\sum_{b\neq a} x^i_b(t) = \sum_b W_{a,b} x^i_b(t)\,
\end{equation}
where $\mu$ is the mutation rate. Solving the linear differential equation, we can compute the probability that a site mutates into a specific new state in the time interval $\Delta t$ as 
\begin{equation}
    p_{\neq} = \frac{1}{A}\left( 1 - e^\frac{A \mu \Delta t }{1 - A} \right)\,,
\end{equation}
while the probability of not mutating is $p_= = 1 - (A-1)p_{\neq}$. Here we set $\Delta t =1$. Hence the probability of evolving from a sequence $\mathbf{v}$ to $\mathbf{v}'$ is 
\begin{equation}
    \pi(\mathbf{v},\mathbf{v}') = p_=^{Nq}p_{\neq}^{N(1-q)} = e^{-N\pot(q)}\,,
\end{equation}
where $q$ is the overlap between the two sequence and $\pot(q) = q\log\frac{p_{\neq}}{p_=} - \log p_{\neq}$. Hence, the probability to go from $\mathbf{v}_0$ to $\mathbf{v}_T$ in $T$ steps is
\begin{equation}
    P(\mathbf{v}_0\to\mathbf{v}_T;T) = \sum_{\{\mathbf{v}\}_{t=1}^{T-1}}  \pi(\mathbf{v}_0,\mathbf{v}_1) \pi(\mathbf{v}_1,\mathbf{v}_2)\dots\pi(\mathbf{v}_{T-1},\mathbf{v}_T) =\sum_{\{\mathbf{v}\}_{t=1}^{T-1}} e^{-N\sum_t \pot(q_t)}\,,
\end{equation}
which can be computed exactly as
\begin{equation}
    P(\mathbf{v}_0\to\mathbf{v}_T;T) = \frac{p_{\neq}^{TN}}{A^N}\left[\left(\frac{p_=}{p_{\neq}}+A-1 \right)^T -\left(\frac{p_=}{p_{\neq}}-1\right)^T \right]^D\left[\left(\frac{p_=}{p_{\neq}}+A-1 \right)^T -(1-A)\left(\frac{p_=}{p_{\neq}}-1\right)^T\right]^{N-D}\,,
\end{equation}
where $D$ is the Hamming distance between the two sequences. The last equation corresponds to Kimura's theory of neutral evolution~\cite{kimuraneutral}. The optimal time for which this probability is maximised, $T^*$, and converges to $1/A^N$ for $T\to\infty$.

\bibliography{ref3,citations}
\newpage

\section{Appendix A: Weights Logo for the WW domain}
 \begin{center}
\includegraphics[page=1,width=\textwidth]{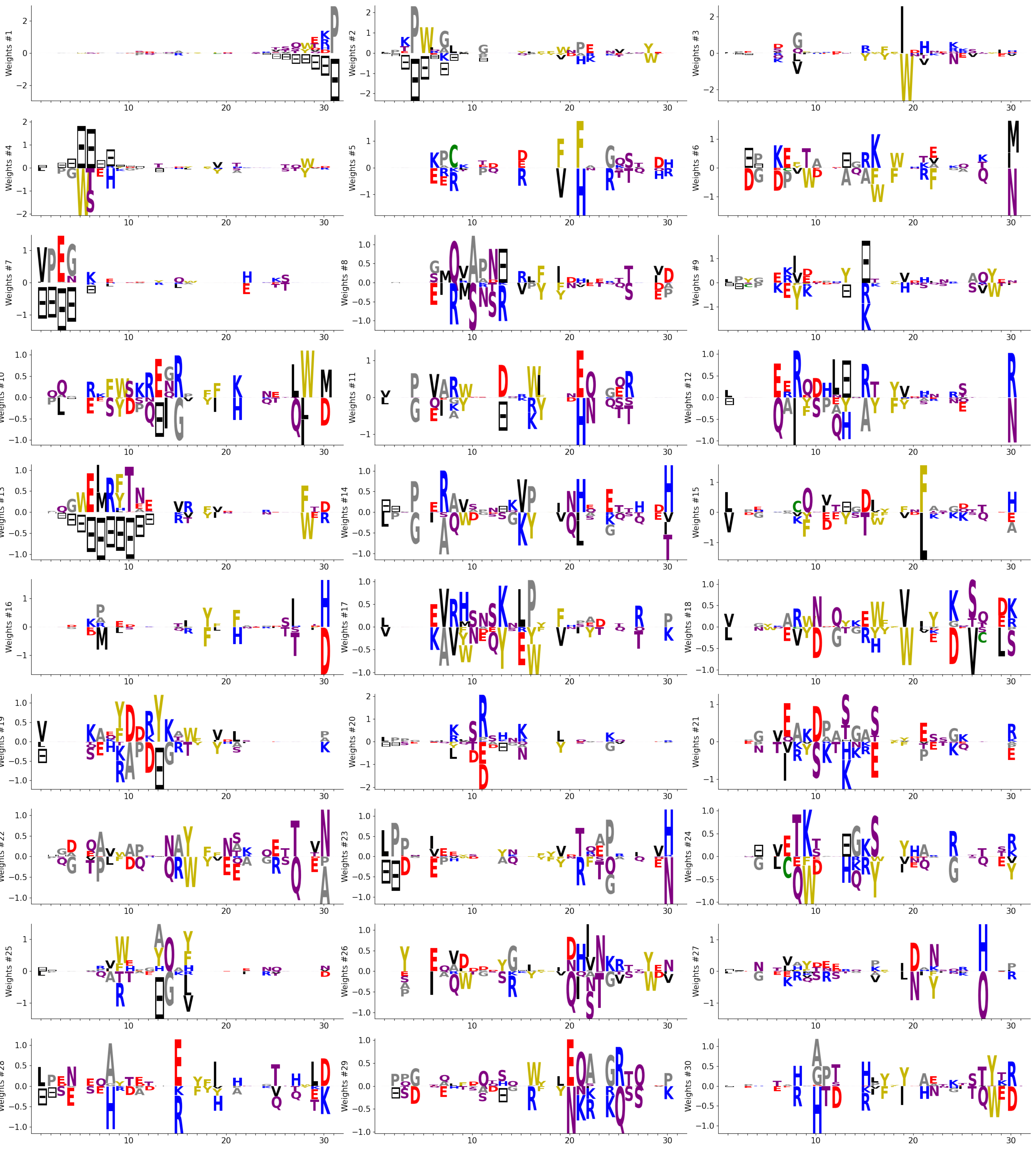}
\end{center}
\begin{center}
\includegraphics[page=2,width=0.9\linewidth,angle=-90]{WW_all_weights.pdf}
\end{center}
\newpage
\section{Appendix B: Weights Logo for the Lattice Proteins}
 \begin{center}
\includegraphics[page=1,width=0.9\linewidth]{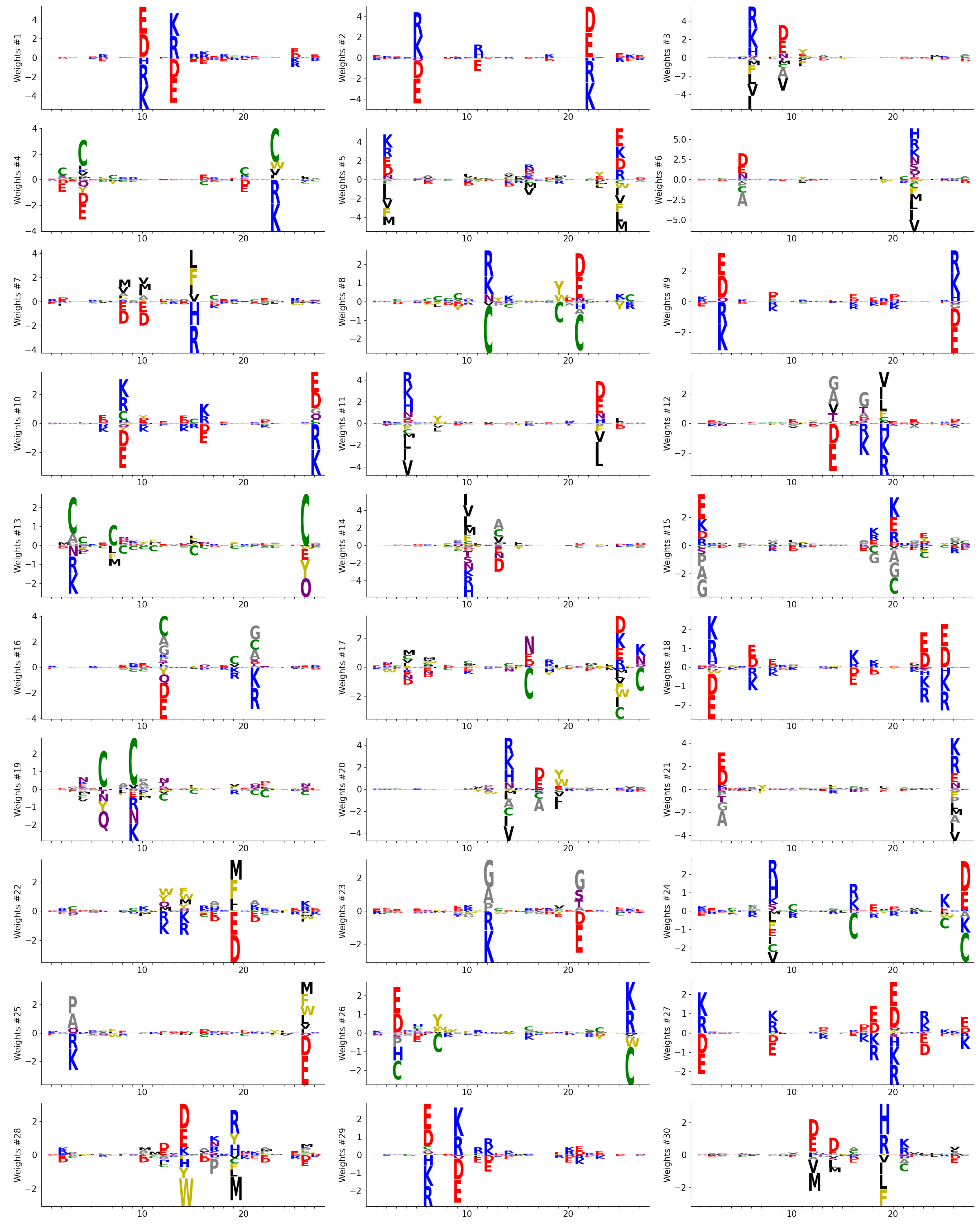}
\end{center}
\begin{center}
\includegraphics[page=2,width=0.9\linewidth]{LP_all_weights.pdf}
\end{center}
 \begin{center}
\includegraphics[page=3,width=0.9\linewidth]{LP_all_weights.pdf}
\end{center}
\begin{center}
\includegraphics[page=4,width=0.7\linewidth,angle=-90]{LP_all_weights.pdf}
\end{center}